\documentclass[aps,pra,superscriptaddress,]{revtex4-2} 

\usepackage{enumitem}
\usepackage{float}
\usepackage{amsmath,amssymb}
\usepackage{graphicx}
\usepackage{dcolumn}
\usepackage{bm}
\usepackage[squaren,Gray,cdot]{SIunits}
\usepackage{enumitem}
\usepackage{xcolor}
\usepackage{ulem}
\usepackage{colortbl}
\usepackage{color}
\usepackage[pdftex]{hyperref} %
\hypersetup{colorlinks,linkcolor={blue!80!black},citecolor={blue!80!black}, urlcolor={blue!50!black}}
\usepackage{changepage}
\usepackage{soul}

\definecolor{editcolor}{rgb}{0.90,0.10,0.25}

\begin{document}

\title{On-demand microfluidic droplet pinching and splitting under local confinement gradients}

\author{Margaux Kerdraon}
 \affiliation{Gulliver CNRS UMR 7083, PSL Research University, ESPCI Paris, 10 rue Vauquelin, 75005 Paris, France}
 \author{Albane Th{\'e}ry}
\affiliation{Gulliver CNRS UMR 7083, PSL Research University, ESPCI Paris, 10 rue Vauquelin, 75005 Paris, France}
\affiliation{Department of Mathematics, University of Pennsylvania, 19104 Philadelphia, Pennsylvania, USA}
\author{Marc Pascual}
 \affiliation{Gulliver CNRS UMR 7083, PSL Research University, ESPCI Paris, 10 rue Vauquelin, 75005 Paris, France}
\author{St{\'e}phanie Descroix}
 \affiliation{Institut Curie, CNRS UMR168, Laboratoire Physico Chimie Curie, Institut Pierre-Gilles de Gennes, PSL Research University, 75005 Paris, France}
\author{Marie-Caroline Jullien}
  \email{marie-caroline.jullien@univ-rennes.fr}
  \affiliation{Gulliver CNRS UMR 7083, PSL Research University, ESPCI Paris, 10 rue Vauquelin, 75005 Paris, France}
 \affiliation{Universit\'e Rennes, CNRS, IPR (Institut de Physique de Rennes) - UMR 6251, F-35000 Rennes, France}
\date{\today}
%

\begin{abstract}

We report the pinching of an elongated liquid droplet confined in a rectangular microchannel. A local variation of the channel topography induces the droplet pinching and can lead to its breakup. The modification of the channel topography is either caused by a reversible local dilation of the channel bottom wall or by a confinement gradient irreversibly printed in the channel.
Interestingly, in both cases, a few micrometres of channel height contraction leads to the droplet's pinching in the direction of the channel width. If this pinching is such that the droplet is no longer confined at the level of its deformation, {\it, i.e.} the liquid thread becomes 3D, the droplet splits in two.
By minimizing the surface energy of the droplet under the channel confinement gradient, we can predict its pinching or breakup, if any. The dynamics of the droplet deformation are then captured by a semi-analytical model that relies on the observation that the deformation is self-similar, and assumes that most of the viscous dissipation occurs in the gutters at the corners of the channel. Conversely, we show that this kinetic model can be used to extract the channel's average topographic variation, down to a few micrometres, by measuring the droplet pinching dynamics under a classical optical microscope.\\ 
%
\end{abstract}

\maketitle
%
\section{Introduction}
The microfluidic field has thrived over the last decades with the promise of miniaturizing chemical and biological assays to reduce costs and enhance their throughput \cite{tabeling2005introduction}. Droplet-based microfluidic platforms are a network of micrometric channels in which water-in-oil or oil-in-water droplets are generated and used as biochemical reactors as an alternative to microlitre plates. For such use, controlling the droplet size, monodispersity, and frequency generation is critical, and these are inevitably related to breakup or coalescence events \cite{baroud2010dynamics}. 

Droplet breakup is a basic functionality required in emulsion science for droplet production. The fragmentation mechanism has been extensively studied in the literature. In the late 19th century, Plateau and Rayleigh showed that droplets of controlled volume can be produced as the result of the destabilization of a liquid thread in an infinite medium to minimize its area-to-volume ratio at constant volume  \cite{plateau1873statique,strutt1878instability}. Contrary to droplets in an infinite medium, the geometric confinement of droplets by walls in microfluidic systems tends to stabilize the liquid thread \cite{guillot2007stability,guillot2008stability,cabezas2019stability}, except when Marangoni flows come into play \cite{clerget2023marangoni}. Microfluidics offers the possibility to produce droplets with a high level of control of dispersity and frequency, with a plethora of reported techniques. Droplets can be produced at the junction between the two immiscible phases, either by a passive method (T or Y junctions \cite{garstecki2006formation,menetrier2006droplet,jullien2009droplet,leshansky2012obstructed}, flow focusing devices \cite{anna2003formation,takeuchi2005axisymmetric,dewandre2020microfluidic}, step channels \cite{barkley2015snap,montessori2018elucidating,dangla2013physical} and confinement gradients \cite{dangla2013droplet,keiser2016washing,taccoen2019order}), or assisted by an active method \cite{zhu2017passive,chong2016active} such as electric field \cite{he2010low}, magnetic field \cite{wu2013ferrofluid}, or centrifugal forces \cite{haeberle2007centrifugal}.
\\
Once produced, it is critical to be able to control {\it in situ} their breakup, to produce two daughter droplets from a mother droplet at will, typically for biotechnology or microbiology dedicated systems in which droplets form microreactors \cite{kaminski2016droplet,scheler2019recent,shang2017emerging}. As for their production, techniques to break droplets can also rely on passive or active methods. To cite a few, passive methods are mostly based on geometric modification of a channel, such as T or Y junctions leading to two downstream channels \cite{menetrier2006droplet,jullien2009droplet,zhou2023breaking,cochard2023droplet}, inserting obstacles in the chanel \cite{nishimura2017breakup,li2022geometrical,salkin2013passive}), using hydrodynamic trap with a bypass \cite{bithi2010behavior}, including a post-array in a cavity\cite{masui2023understanding}, or expanding the chanel \cite{kuang2024droplet}. In this paper, we are concerned with breaking droplets actively. Most of the methods in the literature are very similar to the one associated to their production using external forces \cite{agnihotri2025droplet} such as mechanical/pneumatic \cite{choi2010designed,zhu2011controllable,yoon2013active}, acoustic \cite{jung2016demand}, magnetic \cite{wu2015active,shyam2022magnetofluidic} or electrical ones \cite{link2006electric,han2020droplet}. Our work is in a category that uses mechanical action on the droplets by activating a controlled deformation of the channel walls, which may or may not lead to droplet breakup. Interestingly, our approach does not rely on any physical properties of the droplet content (magnetic, electric, etc) and requires only the integration of a heating resistance. In this paper, we study the mechanisms involved in the breakup of droplets to be able to control it.\\


More specifically, in this paper, we investigate the response of an oil droplet subjected to a local confinement gradient in a microfluidic rectangular channel. We focus on droplets surrounded by an aqueous external phase that do not wet the channel wall. We study the effect of the local confinement gradient on static droplets,  i.e., the droplets do not travel in the channel and remain approximately centered under the local confinement gradient. The confinement gradient is applied on demand using a network of heating resistances that locally dilate the bottom silicone layer of the chip. In this {\it active} device, the local temperature increase can lead to a channel deformation, more precisely a constriction, and to interfacial stresses resulting from a surface tension gradient, called Marangoni stress. To capture the mechanisms that prevail during droplet deformation, we also study the fate of a droplet in a passive device, where a similar constriction is irreversibly shaped in the channel and no thermal effects are present. The advantage of the first configuration (active and reversible) compared to the second one (passive and irreversible) is that it can generate a custom deformation. On the other hand, in the second passive device, the droplet pinching is purely mechanical, and the channel shape is well-characterized. 

The parallel study of the two configurations allows us to determine precisely which mechanism, the thickness gradient or the Marangoni effect, is ultimately responsible for the droplet breakup in the active device. In both configurations, we find that the droplet pinches, forming a neck in the observation plane, to counterbalance the increase in its curvature under the local thickness gradient. When the neck forms, the outer phase drains through the four gutters between the drop and the corners of the rectangular channel, by mass conservation. The droplet deformation is thus capillary-driven and mediated by the viscous dissipation of the flow in the gutters. Beyond a critical thickness gradient, droplets end up breaking. In our setups, it is essential to note that there is no need for a forced flow of the external phase to trigger the droplet breakup.

In the following, we first characterize experimentally the pinching and breakup dynamics and show that in both microfluidic devices, a small local variation of the channel height - 10 to 20 \% of the initial channel height - can induce a significant droplet pinching and even lead to its breakup when the confinement gradient is sufficiently large. We then theoretically study the droplet pinching and fate through the combination of a minimal model relying on surface energy minimisation and $Surface~Evolver$ numerical simulations \cite{brakke1992surface}. We also propose a power balance between the viscous dissipation and the capillary driving force to capture the time evolution of the droplet deformation. 
Finally, we demonstrate that, strikingly, this model can also be used to extrapolate topographical defects of a channel from the observation of droplet pinching dynamics.

\section{Material and methods}

Droplets are composed of mineral oil (Sigma Aldrich M5904). The external phase is an aqueous solution of sodium dodecyl sulfate (SDS, Merck) at a concentration of 2.94 g.L$^{-1}$ in deionized water (critical micellar concentration, $c_{cmc} = 1.92$ g.L$^{-1}$). As such, the droplet does not wet the substrate and is fully surrounded by the external phase. We add fluorescein to the solution at 0.44 g.L$^{-1}$ to enhance the contrast between the droplet and the outer phase under fluorescence microscopy.\\
The viscosity of the aqueous solution (outer phase) is $\eta_w = 10^{-3}$ Pa.s and the viscosity of the mineral oil (inner phase) is $\eta_i = 25 \times 10^{-3}$ Pa.s. We measure the interfacial tension between the inner and the outer phases $\gamma = 11.1$ mN.m$^{-1}$ by a pendant drop method (Kruss DSA30) (Appendix \ref{app:pendantdrop}).
\\
 The microfluidic channel consists of poly(dimethylsiloxane) (PDMS) bonded to a glass slide on top of which a 30-micron layer of PDMS has been spin-coated and cured. We use an oxygen plasma (FemtoScience) for the bonding. Water is injected into the microfluidic channel a few minutes after the plasma exposition to ensure the PDMS channels remain hydrophilic during the experiments. 
We generate the oil droplets through a T-junction at the entrance of the microfluidic device. We set the oil and the aqueous phase flow rates with a pressure controller (Fluigent MFCX-Flex) to control the droplet length. We tune the flow rate of the external phase at the entrance and the channel's output to maintain the droplet {\it static} within the channel, {\it i.e.} located centred at the constriction. 

\begin{figure}[ht]
\centering
\includegraphics[scale=0.85]{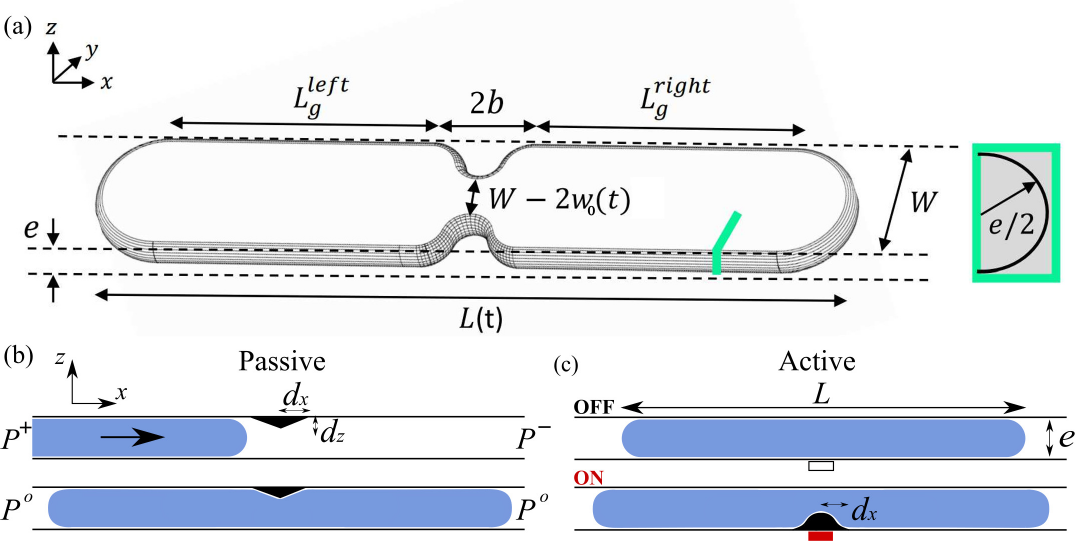}
\caption{ \textit{(a) 3D sketch of the peanut-like deformed droplet confined in the rectangular channel. The green cross-section indicates the gutters' configuration. (b) Side-view of the passive device in which the constriction is already present in the PDMS ceiling (mechanical system). A pressure difference is applied to push and centre the droplet under a constriction of size $2d_x$ along the x-axis and $d_z$ along the z-axis. (c) Side-view of the active device used to induce on-demand the thermal dilation of the PDMS bottom layer (thermomechanical system). The red square indicates the heating resistance. }}
\label{fig:sketch}
\end{figure}

In our experiments, the lengths of the droplet $L$ along the $x$ direction range from $500$ up to $4000$ $\upmu$m. The channel cross-section is rectangular, so the droplet is confined in height and width, see FIG.~\ref{fig:sketch}.a. The ratio between the channel height $e$, along the $z$ axis, and its width $W$, along the $y$ axis, is of the order 1/10, with $e$ varying from $20$ to $40$ $\upmu$m and $W$ varying from $200$ to $400$ $\upmu$m. For clarity, in the following we call \textit{menisci} the curved interfaces of the oil droplet in the (y,z) plane and \textit{gutters} the volume of water trapped between the channel walls and the droplet menisci further down the droplet neck. In such a configuration, the characteristic length of the droplet menisci, as well as that of the gutters cross-section, scales with the channel height $e$, see FIG.~\ref{fig:sketch}a. 
 
We design two types of devices to investigate the droplet's pinching and breakup under a local confinement gradient. In the first device, the channel thickness variation that imposes a confinement gradient on the droplet is irreversibly molded in the chip. By passively pinching the droplet, this device allows us to determine which mechanism prevails in the second device, where a thermally-actuated constriction from a heating resistance triggers the droplet deformation.

\begin{itemize}
    \item \textbf{The passive (irreversible) pinching device}\\
    The confinement gradient is indented in the brass mold with a precision of approximately 100 nm. The indent, characterized by $d_x$ and $d_z$, is observed under a mechanical profilometer to verify that the mold has been successfully milled. As sketched on FIG \ref{fig:sketch}b, the indentation is a ridge with a triangular cross-section. In the experiments, $d_x$ varies from 400 $\upmu$m to 800 $\upmu$m and $d_z$ from 3 $\upmu$m to 10 $\upmu$m. The indentation aspect ratio sets the confinement gradient's tilt angle $\alpha$, $ \tan \alpha = d_z/d_x$, and ranges from $0.2 ^\circ$ to $1.4^\circ$. The PDMS chip is then replicated from the micromilled brass mold. The channel thickness variation that results from the constriction is $e(x)$. In our experiments, we push the droplet towards the confinement gradient using pressure controllers. We set the pressure gradient along the channel to zero as soon as the droplet centre reaches the indentation. The droplet starts to deform as soon as it undergoes the confinement gradient. For this study, we only focus on the period during which the droplet is centred and static under the constriction. This period corresponds to the late times of the droplet deformation (typically the last $40-50\%$ of the full droplet deformation process). 
    \item \textbf{The active (reversible) pinching device}\\
    The active pinching device is inspired by previous works \cite{miralles2015versatile, kerdraon2019self,pascual2019wettability}. In this device, the thermal dilation of the channel PDMS bottom layer above a heating micro-patterned 50 $\upmu$m wide resistance induces the confinement gradient (see FIG.~\ref{fig:sketch}c). Switching on and off the voltage intensity applied to the heating resistance imposes the confinement gradient on the droplet on demand. The extent of the channel thickness variation depends on the voltage intensity applied to the resistance. We characterize its vertical extent $d_z$ using a step microfluidic channel of incremental height, varying from 5 to 10 $\mu$m, see FIG.~\ref{fig:step}a of Appendix~\ref{appendix:calibration}. For temperatures varying from 55 to 110 $\degreecelsius$ at the surface of the resistance, the vertical extent $d_z$ varies between 5 $\upmu$m $\pm$ 1 $\upmu$m  to 10 $\upmu$m $\pm$ 1 $\upmu$m, see FIG. \ref{fig:step}b in Appendix~\ref{appendix:calibration}. The width of the calibration channel is similar to that of the microfluidic device. Consequently, we monitor the height $d_z$ of the confinement gradient by adjusting the voltage across the resistance, while the horizontal extent $d_x$ is unknown. 
    When the voltage is consecutively switched on and off or off and on, the time response of the channel bottom layer dilation is less than 500 ms. As this response is very fast compared to the duration of the droplet pinching, we consider hereafter that the confinement gradient is imposed or cancelled instantaneously in the channel.
\end{itemize}

Note that in our system, the Bond number, which compares the gravity force to the capillary force given by $\rho g e^2/\gamma$, is $\sim 10^{-4}$. We can therefore ignore gravity as capillary forces dominate. Having a cavity thickness gradient on the top or the bottom of the channel makes no difference, and the choice in our devices is solely related to the microfabrication processes. It is easier to manufacture an indentation in the PDMS top wall than the bottom one for the passive device, and it is easier to get a heating resistor on the bottom wall for the active device. 

 In both devices, the confinement gradient is approximated by a ridge with a triangular cross-section, defined by its horizontal and vertical extents $d_x$ and $d_z$, see FIG.~\ref{fig:sketch}b\&c. This hypothesis has been validated by considering other indentation shapes in Appendix \ref{appendix:indentation}. The indentation profile is assumed invariant along the $y$-axis. In the reference frame centred at the indentation, the channel thickness $e(x)$ is a piecewise function:
\begin{equation}
\left\{ \begin{aligned}
\textrm{for}\; -\infty \leq x\leq -d_x, \quad  & e(x)=e\\
\textrm{for}\; -d_x \leq x\leq 0, \quad \quad \, & e(x)=e-d_z-x\,d_z/d_x\\
\textrm{for}\; 0 \leq x\leq d_x,  \quad \quad \quad \; \, & e(x)=e-d_z+x\, d_z/d_x\\
\textrm{for}\; d_x \leq x\leq \infty, \quad \quad \quad & e(x)=e
\end{aligned} \right.
\end{equation}

To quantitatively investigate the droplet pinching, we introduce several parameters in FIG.~\ref{fig:sketch}. The ($x,y$) plane is the observation plane, and ($y,z$) is the out-of-plane one. The time-dependent droplet length during pinching is $L(t)$. The in-plane distance between the channel wall and the droplet interface, which defines the droplet neck profile along the $x$-axis, is defined as $w(x,t)$, see FIG.~\ref{fig:sketch}. The maximum value of this distance, at the centre of the confinement gradient corresponding to $x=0$, is defined as $w_0(t)$. If the deformation is asymmetric, $w_0(t)$ is defined as $=(w_0^1+w_0^2)/2$, see FIG.~\ref{fig:hoTemp}b. Thus, the neck width along the $y$-axis writes $W - 2 w(x,t)$, where $W$ is the channel width, and is defined by $b(t)$ in the $x$-direction. By volume conservation, the neck extension dimensions, $w(x,t)$ and $b(t)$, can be related to the droplet elongation $L(t)$. The droplet deformation is quantified by the spatio-temporal evolution of the neck profile rather than by the droplet length $L(t)$. Finally, eight gutters are located between the droplet meniscus and the microfluidic channel corners, see the green box in FIG.~\ref{fig:sketch}a. We denote $L_g$ the mean gutter length. If the droplet is not perfectly centred under the confinement gradient, we calculate $L_g$ as the equivalent length of the eight gutters placed in parallel next to the droplet neck, $1/L_g=1/L_g^{left}+1/L_g^{right}$. All these notations are illustrated in FIG.~\ref{fig:sketch}. 

In addition, we use the $Surface~Evolver$ \cite{brakke1992surface} software to predict the shapes of droplets reaching an equilibrium when minimizing their surface energy and compare them to our experiments. The full details of the $Surface~Evolver$ simulations are provided in appendix I.


\section{Observations and Results}
\subsection{Experimental observation of the droplet pinching and breakup} 

\begin{figure}[ht]
\centering
\includegraphics[scale=0.75]{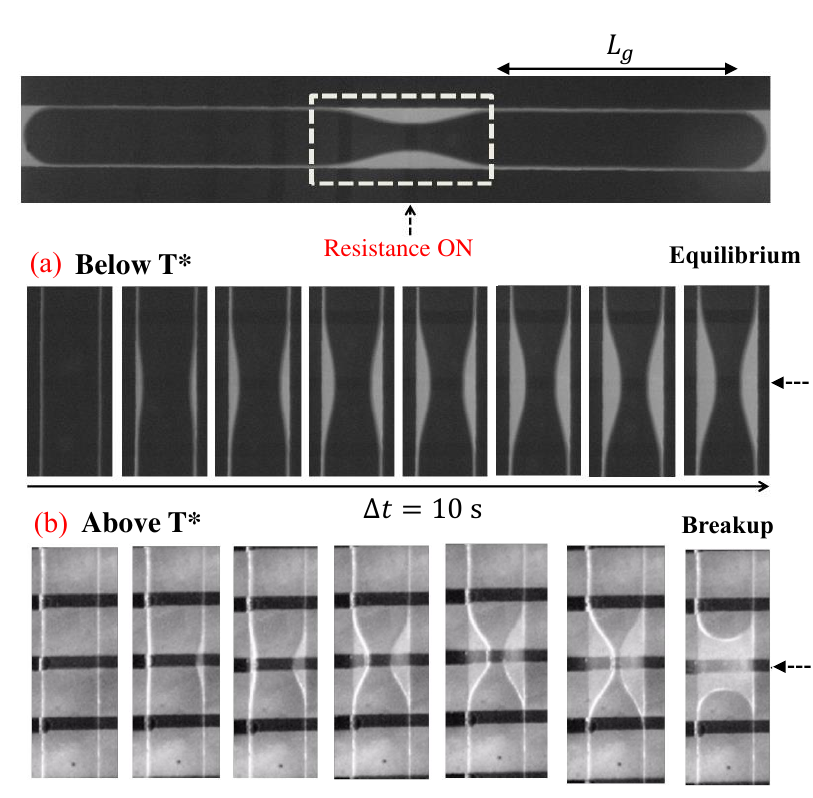}
\caption{ \textit{ Active device, snapshots of the droplet deformation near the heating resistance sketched in dashed arrow. Once a neck forms at the center of the droplet, the number of gutters at the corners of the channel increases from four to eight. The average gutter length is defined as $L_g$.  (a) Droplet pinching at T=55 $\degreecelsius$ ($<T^*$), reaching an equilibrium state. The typical duration of the droplet pinching is a few tens of seconds, (b) Droplet breakup at T=75 $\degreecelsius$ ($>T^*$). }}
\label{fig:illustration}
\end{figure}

When the droplet undergoes the confinement gradient, it pinches in the $(x,y)$ plane like a peanut, so a neck forms in its centre. The droplet deforms for a few seconds to tens of seconds, until it reaches an equilibrium shape or splits into two, see FIG.~\ref{fig:illustration}. 

\begin{figure}[ht]
\centering
\includegraphics[width = \columnwidth]{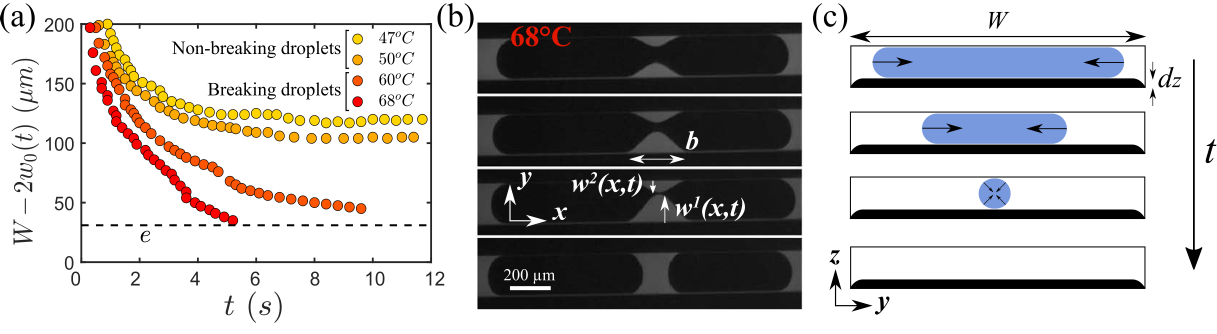}
\caption{ \textit{(a) Time evolution of the neck width, $W-2w_0(t)$, at four different temperatures. At 47$^o$C and 54$^o$C, the neck reaches a constant width after a few seconds, signifying that the droplet has reached an equilibrium shape. At 60$^o$C and 68$^o$C, the neck width decreases until it reaches the confinement height $e$ and the droplets break. In these experiments, the channel dimensions are $e$= 30 $\upmu$m and $W$= 200 $\upmu$ and the droplet initial length is 1300 $\upmu$m. (b) Time-lapse of the droplet deformation and breakup, at 68$^o$C, in the in-plane. In this experiment, the deformation is asymmetric such that $w_0(t)$ is defined as $(w_0^1+w_0^2)/2$. (c) Cross-section sketch of the droplet neck during its pinching at $x=0$. As long as the liquid thread is confined by the channel's top and bottom walls, the droplet does not break, in agreement with Guillot et al., who formalized the stabilisation of a liquid jet by the confinement \cite{guillot2007stability, guillot2008stability}. When the liquid thread is no longer confined, \textit{i.e.} $W-2w_0=e$, the neck is not at equilibrium and the droplet breaks into two.   }}
\label{fig:hoTemp}
\end{figure}

\begin{figure}
\centering
\includegraphics[scale=0.67]{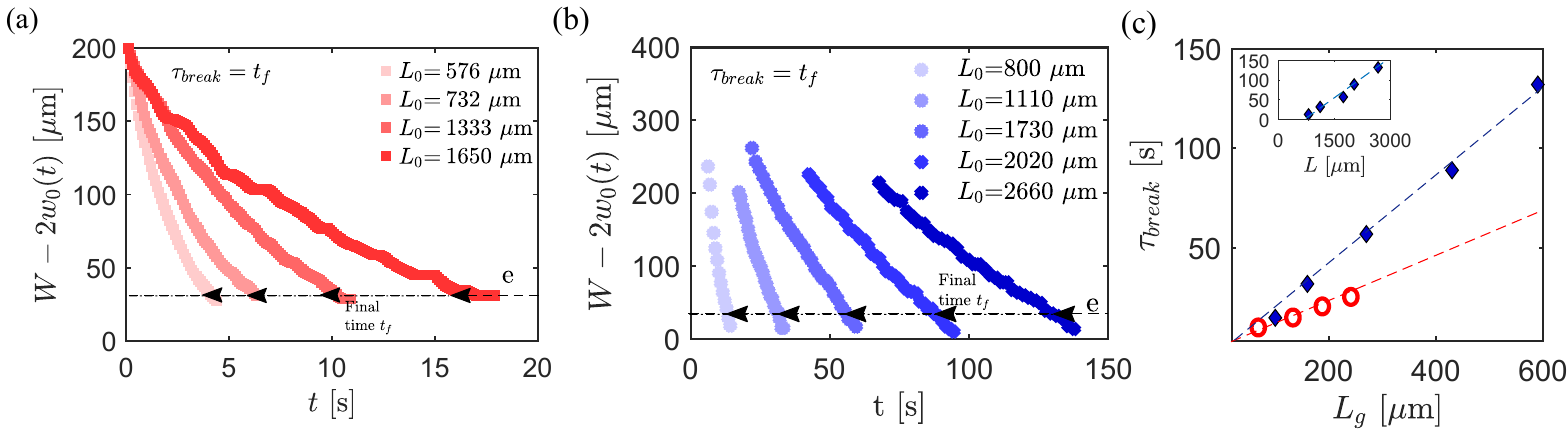}
\caption{\textit{ Time evolution of the neck width $W-2w_0(t)$, for several initial droplet length $L$, in both (a) a thermomechanical  ($e$= 30 $\upmu$m, $W$= 200 $\upmu$m and $T=60^oC$) and (b) a mechanical  ($e=30$ $\upmu$m, $W=400$ $\upmu$m, $d_x=400~\upmu$m, and $d_z=5$ $\upmu$m) system. In both cases, droplets eventually break regardless of their initial length. 
(c) Evolution of the breakup time versus the equivalent length of the gutters at the four corners of the channel, for the passive device (blue) and the active device (red). The break-up time tends to zero when the gutters' length is zero; it is longer in the 400 µm large passive device (blue) as compared to the 200 µm large active device (red), as the volume to fill from the gutters is larger. In the inset, we plot the breakup time as a function of the initial droplet length.}}
\label{fig:taubreak}
\end{figure}

In the active pinching device, the droplet's fate depends on the maximum temperature increase in the channel. Beyond a certain temperature value $T^*$, the droplet deforms until its neck width $W-2w_0(t)$ gets smaller than the channel height $e$, \textit{i.e.} the droplet neck becomes a three-dimensional thread and then breaks. Conversely, below $T^*$, the droplet deforms until it reaches an equilibrium shape in which the neck width is larger than the channel height, \textit{i.e.} the droplet remains confined by the top and bottom walls, see FIG.~\ref{fig:illustration} and FIG.~\ref{fig:hoTemp}a \& b. The temperature threshold $T^*$ delimits the breakup from the non-breakup regime. Figure \ref{fig:hoTemp}(a) shows the temporal evolution of the neck width at four different temperatures. We see that, at 47$^o$C and 50$^o$C, the droplet reaches an equilibrium shape while, at 60$^o$C and 68$^o$C, the droplet ends up breaking when its diameter is comparable to the channel height $e$, see FIG.~\ref{fig:hoTemp}a \& c. At that time, the liquid thread forming the neck switches from a 2D confined shape to a 3D shape where the collapse is always unstable \cite{dollet2008role,van2011microbubble}. In the experiment illustrated in FIG.~\ref{fig:hoTemp}, $T^*$ ranges between 50 and 60 $\degreecelsius$. While the breakup of the droplet depends on the temperature increase in the vicinity of the resistance, {\it i.e.} depends on the wall deformation, the dynamics of its deformation depends on the initial length of the droplet. The longer the droplet, the more time it takes to deform, see FIG.\ref{fig:taubreak}a\& c. 

In the passive pinching device, the fate of the droplet depends on the geometric configuration of the channel. More precisely, the droplet breakup criterion depends on the extent of the channel thickness gradient, defined by $d_x$ and $d_z$ and the other geometric parameters for the channel. As in the active system, the pinching dynamic of the droplet depends on its length, see FIG.\ref{fig:taubreak} b \& c, while the breakup criterion does not.

The break-up time is defined as $t_f$. In the active device, the confinement gradient is applied when the droplet is approximately centered on top of the resistance. However, in the passive device, the droplet starts to deform as soon as it encounters the constriction. In this device, we only conduct the temporal analysis of the droplet deformation when gutters are well defined on each side of the droplet, along the $x$-axis, see FIG.\ref{fig:taubreak}b. Interestingly, in both devices, the final droplet breakup time $t_f$ varies linearly with the droplet initial length $L$ and the gutter length $L_g$. The break-up time tends to zero when gutters disappear ($L_g=0$), see FIG.\ref{fig:taubreak}c.


\subsection{Description of the mechanisms for droplet deformation}

\begin{figure}[ht]
\centering
\includegraphics[scale=0.6]{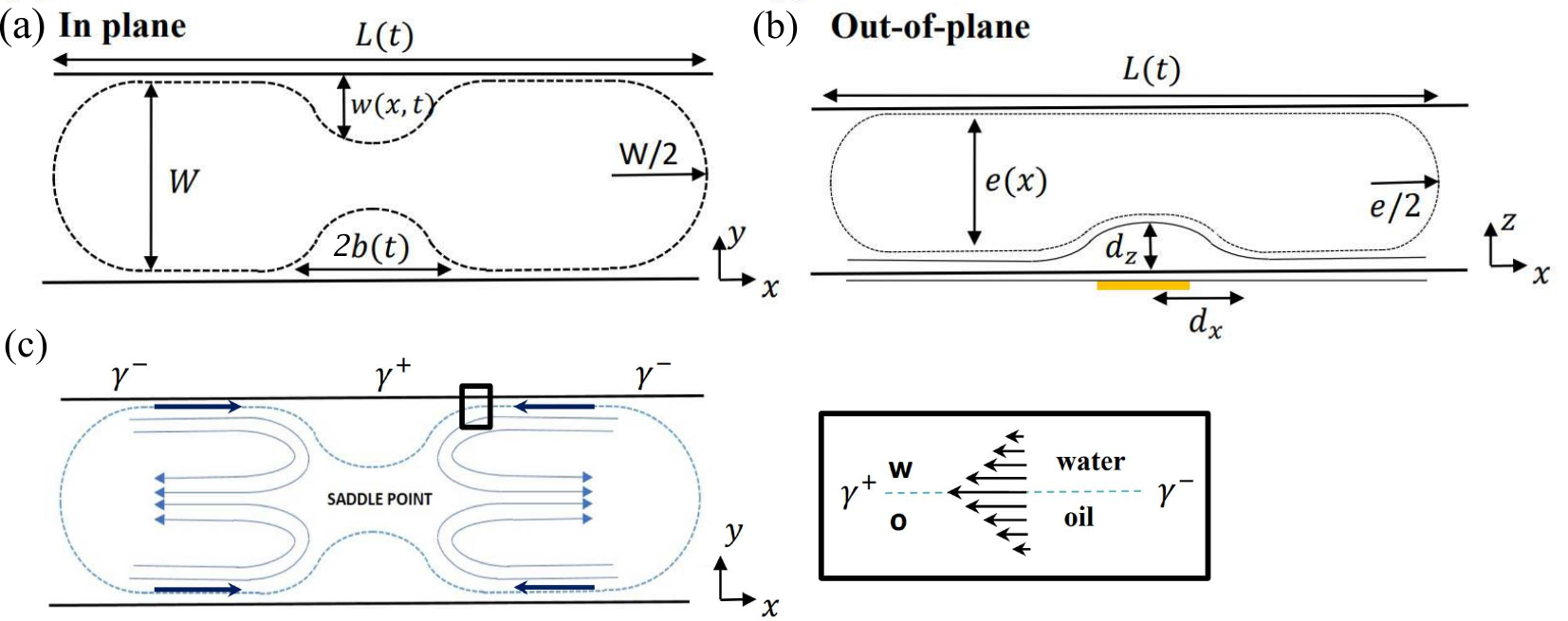}
\caption{\textit{(a) Sketch of the pinching droplet in the in-plane view, \textit{i.e.} the observation plane. The droplet deformation is defined by its neck profile $w(x,t)$, its width $b(t)$, and the droplet length $L(t)$ along the x-axis. (b) Sketch of the droplet deformation in the out-of-plane direction, perpendicular to the observation plane. Here, the droplet deforms because of the thermal dilation of the silicone sub-layer of the microfluidic chip (active device), the dimensions of which are $d_x$ and $d_y$ along the x-axis and the y-axis, respectively. (c) Flows governed by Marangoni thermocapillary gradients at the oil/water interface. }}
\label{fig:sketch2}
\end{figure}

In the active system, two mechanisms, induced by the local temperature increase, can be responsible for the droplet deformation. The first is a mechanical effect associated with Laplace pressure gradients, and the second is a Marangoni effect associated with surface tension gradients, see FIG.~\ref{fig:sketch2}. The passive device, involving solely a mechanical forcing, allows us to discriminate between both effects. 

\vspace{6 pt}

\textbf{1/ Mechanical effect}\\
Given that the radius of the droplet meniscus equals $e/2$ in the out-of-plane dimensions, the channel deformation described by $e(x)$ leads to a decrease of the local droplet meniscus radius $e(x)/2$, see FIG.\ref{fig:sketch}c. The droplet's mean curvature is 
$
\mathcal{C}=\displaystyle\frac{\partial_x^2 w}{\left(1+(\partial_x w)^2 \right)^{3/2}}+\frac{2}{e(x)}
$, where $\displaystyle\frac{\partial_x^2 w}{\left(1+(\partial_x w)^2 \right)^{3/2}}$ is the in-plane curvature~\cite{dangla2013droplet}. At equilibrium, the Laplace equation sets the curvature to be constant. Therefore, an increase in the droplet out-of-plane curvature, induced by the indentation, must be compensated by a decrease in the in-plane one.  
At $t=0$, the in-plane curvature of the droplet is zero. The constriction induced by the thermal dilation of the channel in the active device, or by a permanent obstacle in the passive device, is hence counterbalanced by the formation of a neck, the curvature of which is negative, as described by Dangla $et.$ $al.$ \cite{dangla2013droplet,dangla2013physical}. A rough estimate of the curvature-driven normal stress gives $\Delta P\sim 10^2$ Pa.
By mass conservation, the outer phase (water) drains towards the centre of the droplet through the gutters to occupy the volume released by the formation of this neck. In the meantime, the volume of oil is conserved such that the droplet extremities move slowly away from the constriction. During the process, the droplet exhibits a characteristic peanut-like shape, as in FIG.~\ref{fig:illustration}. 
\vspace{6 pt}

\textbf{2/ Thermal Marangoni effect (for the active device only)}\\
A surface tension gradient can stem from the local temperature increase in the channel at the level of the heating resistance, see FIG.\ref{fig:sketch2}c. In our experimental setup, the surface tension increase with temperature is equal to $\partial_T \gamma \sim 2.3\times 10^{-5}$ N.m$^{-1}$.K$^{-1}$ while the temperature gradient is typically equal to $\partial_x T=10^{4}$ K.m$^{-1}$ \cite{kerdraon2019self}. The Marangoni stress $\partial_x \gamma= \partial_T \gamma \partial_x T $ that results from these two gradients is thus of the order of $ 2\times 10^{-1}$ N.m$^{-2}$, which is low as compared to the curvature-driven stress.
Being oriented from low to high surface tension regions, Marangoni flows are directed towards the droplet central region where the neck forms. By mass conservation, a saddle point should form in this region, leading the droplet to pinch and eventually break. However, the droplet's ability to sustain an equilibrium shape below $T^*$ despite the temperature gradient strongly suggests that thermal Marangoni effects are negligible in this study. 
The hypothesis that only mechanical effects drive the deformation of the droplet is further validated by comparing the active device to the passive one, in which the droplet also displays a steady-state deformation for low indentations and breaks up above a critical threshold.

In a nutshell, the confinement gradient must reach a critical value to trigger the droplet breakup in both the mechanical (passive) and the thermomechanical (active) devices. Conversely, if this gradient is not sufficient, the droplet deforms until it reaches an equilibrium shape, where the horizontal curvature balances the vertical curvature induced by the deformation, see FIG.~\ref{fig:illustration} \&~\ref{fig:hoTemp}.

\begin{figure}[t]
\includegraphics[scale=0.515]{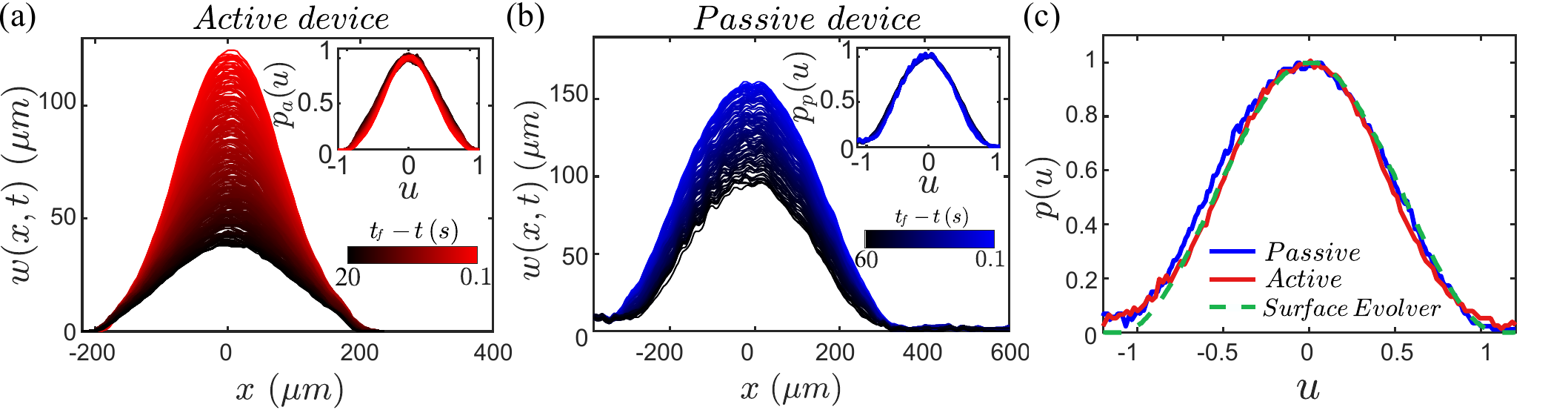}
\caption{\textit{  (a) Time evolution of the neck profile $w(x,t)$ for an active device ($W= 400$ $\upmu$m, $e= 30$ $\upmu$m and $T=60^\circ$C). The droplet is
1600 $\upmu$m long. Inset: Collapse of the rescaled profile $p_a=w/w_0$ versus $u=x/b$, with $t_f-t$ the time remaining before breakup. (b) Time evolution of the neck profile $w(x,t)$ for a passive device ($d_z$ = 5 $\upmu$m, $d_x$ = 400 $\upmu$m, $e$ = 30 $\upmu$m and $W$ = 400 $\upmu$m). The droplet is
1600 $\upmu$m long. Inset: Collapse of the rescaled profile $p_p=w/w_0$ versus $u=x/b$. (c) Comparison between the self-similar functions of the $Surface~Evolver$ simulations, the active and passive devices, $p_a$ 
 and $p_p$, respectively. The collapse suggests that the same mechanism drives the breakup in both systems, ruling out the contribution of Marangoni effects.}}
\label{fig:selfsimilar}
\end{figure}

We track the droplet pinching dynamic in both devices by measuring the time evolution of the neck profile along the x-axis defined by $w(x,t)$, see Figure \ref{fig:selfsimilar}. Interestingly, the neck extension along the $x$-axis, defined by $b(t)$, is constant over time, and will be denoted $b$ in the following.
By judiciously rescaling the spatiotemporal evolution of the droplet deformation in both devices, we find that it is self-similar over time \cite{mcgraw2012self,kerdraon2019self}, see FIG.\ref{fig:selfsimilar}. The neck profile is thus described by a self-similar function of the form $w(x,t)=w_0(t)p(u)$, with $u=x/b$ and $b$ constant. This rescaling is the signature of a quasi-static deformation that can be understood considering that the time scale of the viscous dissipation in the neck is much smaller than in the gutters. As a result, for any given time, the neck region is in a local equilibrium, and the deformation is quasi-static. The insets in FIG.~\ref{fig:selfsimilar}a \&b display the functions $p_a$ and $p_p$ for the active and the passive devices, respectively. 
These functions collapse on the profile obtained from simulations of the equilibrium shapes of a droplet placed in similarly constrained geometries using the \textit{Surface Evolver} software, see FIG.~\ref{fig:selfsimilar}c and Appendix~\ref{appendix:codeSE}.
The good agreement between the experimental profiles, $p_a$ and $p_b$, and the one simulated \textit{in-silico} confirms the hypothesis that the droplet is mainly deformed by the confinement gradient imposed in the channel. The matching of the profiles underlines that the neck shape results from a constant curvature along the droplet, the negative curvature compensating the additional Laplace pressure above the indentation. The other mechanisms, such as the thermal Marangoni effect, that can come into play in the active device, will be disregarded subsequently, even at such high temperature gradients ($\partial_x T\sim$ 10$^4$ K.m$^{-1}$). 
Once we determined that we could neglect the thermocapillary effect, we probed a range of larger drops using a mechanical device.

The droplet breakup occurs when the neck starts to be 3D,  {\it i.e.} when $W-2w_0(t)\sim e$, in agreement with previous studies \cite{dollet2008role,van2011microbubble}. 

\subsection{Interpretation and model for the final shape of the droplet} 
In this section, we propose a simple analytical model to predict the final equilibrium shape of the droplet or its breakup, given the geometric constraints of the channel.
Consistently with our above analysis of the driving mechanisms, our model assumes that the surface energy minimization under the local confinement gradient drives the droplet deformation. 

The following criterion sets the final state of a droplet undergoing a mechanical deformation:
\begin{itemize}[noitemsep,topsep = 0em]
    \item the droplet reaches a stable, peanut-like shape when the neck is thicker than the channel height, $W-2 w_o(t_{eq})>e$ with $t_{eq}$ is the time at which the system reaches equilibrium, and
    \item the droplet breaks when the neck becomes thinner than the channel height at any time $t$, $W-2w_o(t) \leq e$. 
\end{itemize}
Note that our studies, both experimental and theoretical, are based on confined drops that can generate gutters, i.e. of length $L\geq 2b$ to adopt a peanut shape, a criterion that guarantees the existence of gutters. For smaller drops, Figure \ref{fig:taubreak}c shows that the breakup time tends to 0 when $L_g$ tends to 0, which is, of course, unphysical. In this regime, viscous dissipation actually no longer takes place in gutters located in corners but over the entire height of the channel, significantly reducing viscous dissipation and hence the breakup time. This regime is out of the scope of the paper.
In the case of our elongated drops ($L\gg W$), we consider that the droplet length $L$ is constant over time, see Appendix G.2. for more details.
The surface energy of the droplet writes $E = \gamma S$ where $S$ is its surface area and the surface tension $\gamma$ is assumed to be constant over time. 
The excess surface of the droplet, defined as $\Delta S (t)$, corresponds to the difference between the surface area of the droplet 
subjected to the confinement gradient (peanut-like shape) and its initial surface area in the undeformed channel (plug-like shape).
We compute $\Delta S (t)$ as a function of the local variation of the channel height $e(x)$, relying on mass conservation to obtain the variation of the droplet length $\Delta L$, see Appendix~\ref{appendix:staticmod} for more details. At any given time $t$, the total surface variation of the droplet writes
\begin{equation}
\Delta S (t) = 4 \int^{{b}}_{0} \left[\pi e(x)\left(\sqrt{1+\partial_x w^2}-1 \right) \left( 1 + \frac{\lambda}{4} \frac{e(x)}{e} \right)   -   w(x,t) \left( 1+   \sqrt{1 + \left(\partial _x e \right)^2}  + 2 \lambda \frac{e(x)}{e} \right) \right] \textrm{d} x  
\label{eq:deltaS}
\end{equation}
where $ \lambda = \frac{W + \left(\frac{\pi}{2}-1 \right)e }{W + \left(\frac{\pi}{4}-1 \right)e}$
is a geometric parameter of the unindented channel that verifies $1 < \lambda < 2$. Despite $ \sqrt{1 + \left(\partial _x e \right)^2}\sim 1$, this term is formally kept as small deformations of the wall lead to large deformation of the droplet, up to their breakup.

We now use our finding that the neck profile is self-similar, $w(x,t)=w_0(t) p(x/b)$, to write $\Delta S$ as a function of 6 geometric parameters, 
\begin{itemize}[label=--, itemsep=0em, topsep=0em]
    \item the neck extension, defined by its minimal width $W - 2 w_0(t)$ and its extension $b$
    \item the channel size, with width $W$ and height $e$ and
    \item the dimensions of the triangular constriction, $d_x$ and $d_z$, 
\end{itemize} 
and the dimensionless function $p$ and its derivative $p'$, obtained from a polynomial fit to the experimental and numerical profiles, see FIG.~\ref{fig:selfsimilar}. Details are provided in Appendix~\ref{appendix:pfunction}. 

\begin{figure}[t]
\centering
\includegraphics[scale=0.53]{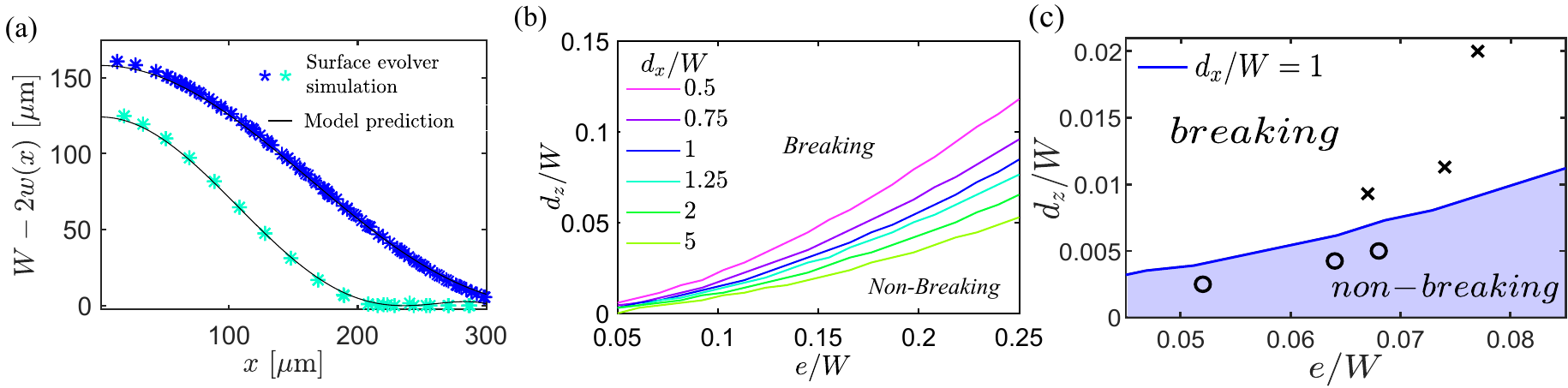}
\caption{ \textit{(a) Comparison between a droplet  profile predicted by \textit{Surface Evolver} simulations and by our model, for two sets of channel geometries ($d_z$ = 4 $\upmu$m, $d_x$ = 400 $\upmu$m, $e$ = 35 $\upmu$m and $W$ = 400 $\upmu$m (blue) and
$d_z$ = 9 $\upmu$m, $d_x$ = 300 $\upmu$m, $e$ = 40 $\upmu$m and $W$ = 200 $\upmu$m (green)). (b) Breakup criteria derived from our calculations for different sets of geometric parameters ($e$, $W$, $d_z$, $d_x$). The parameters are normalized by the channel width $W$. The curves correspond to the spectrum of equations $W-2w_0(t_f)=e$, calculated for several ratios $d_x/W$. (c) Comparison between the theoretical predictions and the experiments for six mechanical devices, with $W$ varying from 200 $\upmu$m to 400 $\upmu$m, $e$ varying from 20 $\upmu$m to 40 $\upmu$m, $d_x$ from 400 $\upmu$m to 800 $\upmu$m and $d_z$ from 3 $\upmu$m to 10 $\upmu$m. Each dot corresponds to a given indentation of the mechanical device and has been tested on droplet lengths ranging from $500$ up to $3000 \mu$m, showing that the droplet fate does not depend on its length.}}
\label{fig:diag}
\end{figure}

By numerically minimizing the excess surface of the droplet $\Delta S$, we then predict the equilibrium shape of the droplet, characterised by the equilibrium dimensions of the neck, $W-2w_0(t_{eq})$ and $b$.
We then validate the predictions of $b$ and $w_0(t_{eq})$ from our semi-analytical model using $Surface~Elvover$ simulations and find an excellent agreement for different channel geometries, see FIG.~\ref{fig:diag}a, for non-breaking droplets. Note that this equilibrium shape does not depend on the droplet length but solely on the different geometric parameters (channel geometry and indentation).

Using the above breakup criterion, $W-w_0(t_{eq})<e$, we now can build a phase diagram based on the dimensionless parameters, $d_z/W$, $d_x/W$ and $e/W$, to predict the fate of the droplet in a wide range of channel geometries, see FIG.~\ref{fig:diag}b.
We compare these predictions to the results of our mechanical passive devices, for which we have access to the exact dimensions of the deformation. All the passive devices designed for this study perform in agreement with our predictions, see FIG.~\ref{fig:diag}c. Interestingly, Fig. 7 shows that the fate of a droplet depends intricately on all geometric parameters, not solely on the confinement gradient.

For the active thermomechanical device, we do not have access to the horizontal extent $d_x$ of the channel thermal dilation with sufficient prediction to assess the performance of this equilibrium, steady prediction. However, we now show that coupling these steady predictions to a dynamic model for the deformation allows us to quantify the droplet behaviour in both devices and recover the indentation size in active devices.

\subsection{Dynamics of the droplet deformation} 

Droplet pinching is driven by surface energy minimisation and limited by viscous dissipation in the droplet oil and the external aqueous phases. The linear increase of the breakup time $\tau_{break}$ with the gutters' length $L_g$ suggests that the viscous dissipation is mainly located in the gutters, even though the viscosity of the outer phase is smaller than the viscosity of the inner phase ($\eta_i/\eta_w=25$). Moreover, it is interesting to note that the time evolution of the droplet pinching, displayed in Figure~\ref{fig:taubreak}a \& b, is radically different from the typical time power-law that governs the evolution of unstable non-confined liquid threads \cite{day1998self}. A simple scaling analysis for droplets reaching equilibrium predicts an exponential relaxation to the deformed shape, but breaks down for larger deformations and breaking droplets (Appendix \ref{appendix:dynamicmod}.1). We therefore build a dynamical model for the dynamics of equilibrium and breaking droplets, relying on the empirical observation that the width of the deformation is independent of time, $b(t) = b_{eq}$ as given by our static model, and considering the self-similarity of the neck profile.

\begin{figure}[t]
\centering
\includegraphics[scale=0.75]{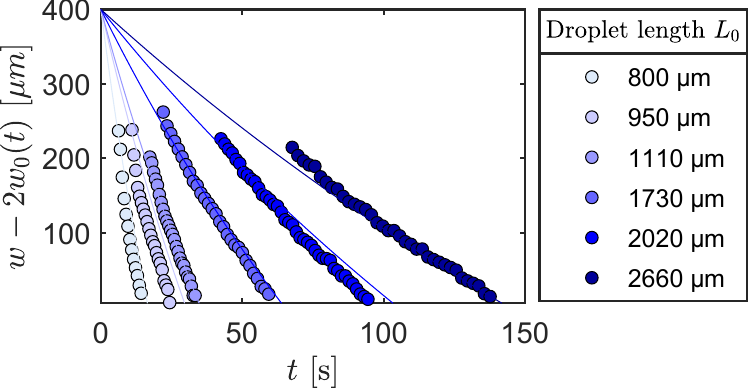}
\caption{ \textit{Time evolution of the neck width $W-2w_0(t)$, for several initial droplet length $L$, in a mechanical device  ($d_x=400~\upmu$m, $d_z=5$ $\upmu$m, $e=30$ $\upmu$m and $W=400$ $\upmu$m). We fit the experimental data with our predictive model ($K=4.68\,10^{3}$).  }}
\label{fig:model}
\end{figure}

To model the droplet pinching dynamic, we write a power balance that compares the viscous dissipation power $\mathcal{P}_g$ with the droplet surface energy increase per unit of time, $\mathcal{P}_\Sigma = \gamma\partial_t (\Delta S)$, with $\Delta S$ given in eq.~\eqref{eq:deltaS}.
 The rate of change in the droplet surface area can be written as $\partial_t (\Delta S)=\partial_t w_0 \mathcal{J}(w_0,b)$, with $\mathcal{J}$ an integral that has the dimension of a length and depends on the geometric parameters of the neck and of the channel. 
 
On the other hand, the viscous dissipation scales as $\mathcal{P}_g \sim \eta_w \displaystyle\frac{v^2}{e^2}e^2L_g$, where $v$ is the typical velocity of the outer phase in the gutters, $\eta_w v/e$ is the order of magnitude of the shear stress in the gutters and $e^2L_g$ is the approximate estimation of the gutter volume. The flow velocity in the gutters $v$ can be assessed by mass conservation, $v \sim \frac{1}{e^2} \partial_t \Omega_3$, with $\Omega_3$ the volume of water entering the neck region as it forms (see Appendix~\ref{appendix:staticmod}). 
As a result, the viscous dissipation power can be written as $\mathcal{P}_g=K \displaystyle  \frac{\eta_w L_g}{e^4} \mathcal{I}(w_0,b)(\partial_t w_0)^2$, where $\mathcal{I}$ is an integral function that depends on $w_0$ and on the channel geometry and has the dimension of a surface squared. 
The dimensionless prefactor $K$ stands for a permeability constant that accounts for the complexity of the flow in the gutters, including a no-slip velocity at the walls, the tangential stress continuity at the fluid/fluid interface, and the complex geometry of the gutter. 

Finally, the power balance allows us to write the following differential equation that governs the time evolution of the neck extension $w_0(t)$ given the constant neck extension $b_{eq}$ from our static model (see Appendix \ref{appendix:dynamicmod}.2 for the detailed calculation),
\begin{equation}
    \partial_t w_o = K \displaystyle\frac{\eta_w L_g}{ \gamma e^4} \frac{\mathcal{I}(w_0,b_{eq})}{\mathcal{J}(w_0,b_{eq})}. 
\label{eq:powerbalance}
\end{equation}
Equation \eqref{eq:powerbalance} is solved numerically, and the adjustable parameter $K$ is set by fitting the theoretical time evolution of the neck width $W-2w_0(t)$ to the experimental
data from the passive device, for which the geometric parameters $d_x$, $d_z$ , $W$ and $e$ are well characterized, see FIG.\ref{fig:model}.
We find $K=4.68\times 10^{3}$, independently of the droplet length. We run \textsc{Comsol} simulations in order to estimate $K$ in the gutters using two limits of the interface boundary conditions at the liquid-liquid interface, either a no-slip or a stress-free one, and found respectively $K = 6 \times 10^3$ and $K = 2.8 \times 10^3$, underlining the relevance of our approach, see appendix \ref{appendix:flowgutter}.

We identify two limit regimes for the deformation dynamics, consistent with our experimental observations. In the non-breaking regime when the equilibrium deformation is small, the system can be approximated by a scaling law for the curvature-driven deformation at the neck (Appendix~\ref{appendix:dynamicmod}1). The above deformation rate $\partial_t w_0$ scales linearly with the neck size, and the neck relaxes exponentially to its equilibrium shape, consistently with FIG.~\ref{fig:hoTemp}a. 
On the other hand, in the limit where the static deformation is larger than the channel width ($w_{0,eq} \gg W$) and the indentation triggers a fast breakup, the rate of change in the neck deformation becomes independent of the neck size (see Appendix~\ref{appendix:dynamicmod}2). As a result, the deformation grows linearly up to the breaking point as in FIG.~\ref{fig:taubreak}b. 

\begin{figure}[t]
\centering
\includegraphics[scale=0.63]{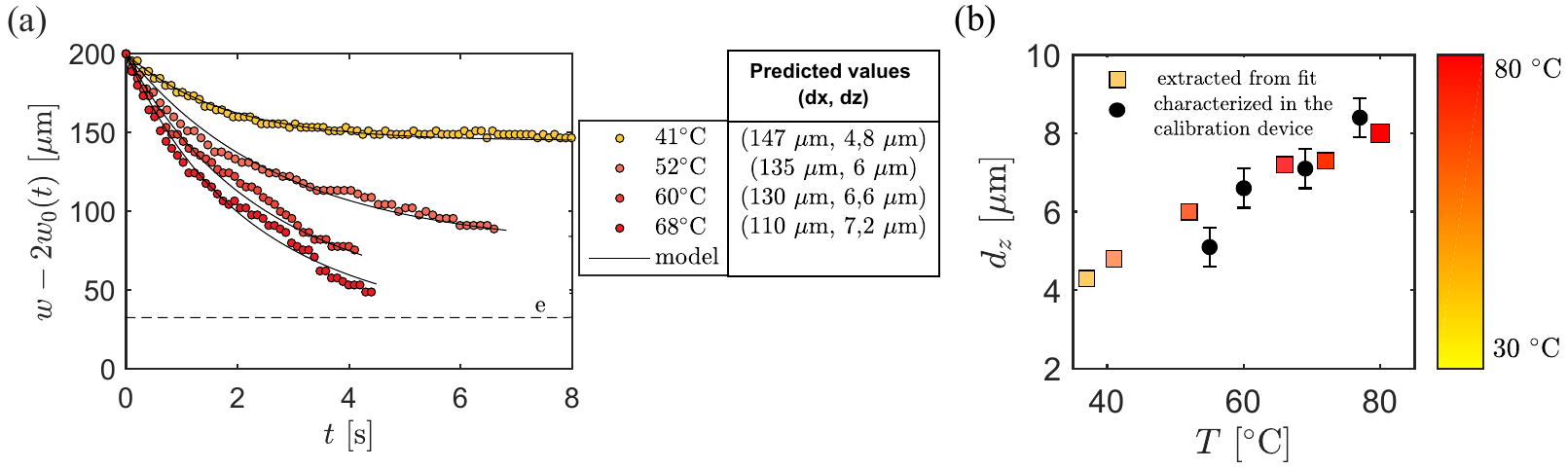}
\caption{\textit{(a) Fit of the experimental profile of the neck width ($W-2w_0(t)$) by our model, in order to determine the channel constriction dimensions $(d_x,d_z)$ of an active device ($e$= 30 $\upmu$m and $W$= 200 $\upmu$m), at several temperatures, for a 1300 µm long droplet. (b) Comparison between the $d_z$ values extracted from the fit and the $d_z$ values extracted from the calibration device, see Appendix~\ref{appendix:calibration}.}}
\label{fig:modelinverse}
\end{figure}

Strikingly, by setting the permeability $K$ to the same value $K=4.68\times 10^{3}$, the model conversely allows us to probe the channel topography in the active device. 
As a reminder, in this device, the vertical extent of the constriction $d_z$ has been characterized as a function of the maximum temperature increase in the channel, while its horizontal extent $d_x$ remains unknown. To determine $d_x$ from the above model, we define the dimensionless function $N$ that calculates the difference between the theoretical and the experimental time evolution of the neck profile, $w_0^{th}$ and $w_0^{exp}$, respectively. $N$ is defined as $N(d_x,d_z)=4[w_0^{exp}(t)-w_0^{th}(t)]^2/e_{rr}^2$ with $e_{rr}$ being the typical error of the image processing, which is considered to be twice the pixel size, approximately equal to $10\,\upmu$m.
The extent of the topographic variation induced by the thermal dilation of the channel walls in the active device, $d_x$ and $d_z$, is calculated by minimizing the function $N$, assuming that the dilation can be approximated to a constriction with a triangular cross-section, with this assumption discussed in Appendix~\ref{appendix:indentation}. Figure 14 of the Appendix \ref{appendix:nsensitivity} shows the typical outputs of such a minimization procedure, showing an important sensitivity of function $N$ on the parameters $d_x$ and $d_Z$. The prediction of $d_x$ and $d_z$, for four sets of experimental data, is illustrated in FIG.\ref{fig:modelinverse}a. The predicted values for $d_z$ are in good agreement with the  $d_z$ values that were characterized using the calibration device described in Appendix~\ref{appendix:calibration}, see FIG.\ref{fig:modelinverse}b. Unfortunately, we cannot verify the   $d_x$ values since they have not been properly characterized. Surprisingly, they seem to decrease with temperature, which might be related to the mechanical response of the silicone layer to high local thermal stresses. Such a response to high thermomechanical stress is out of the scope of this paper. 
Moreover, it is interesting to note that the sensitivity of $N$ to small variations in $d_x$ and $d_z$ is high, indicating that the model should be able to detect out-of-plane defects on the order of hundreds of nanometers, which are tens of micrometres large in the in-plane, see Appendix~\ref{appendix:nsensitivity}. 
More generally, we believe that observing the time evolution of in-plane droplet pinching can reveal small topographic defects in microfluidic channels, which are difficult to measure with optical techniques but can be quantified using our model.

%
\section{Conclusion}
In this paper, we present a novel thermomechanical system that allows for an in situ breakup of droplets in microfluidic rectangular channels due to local variation of the channel topography.
This topography change is induced by micropatterned resistances that locally heat and expand the silicone layer coated on the glass slide of the microfluidic chip.
Strikingly, we find that even small deformations in the channel geometry can lead to large reversible deformations of the droplet or a breakup, providing a controlled and practical mechanism for selectively breaking a droplet after it has been formed.

We study the precise mechanism for droplet breakup experimentally and theoretically and quantify the threshold between reversible deformation and breakup.
We investigate the behaviour of a non-wetting droplet confined in a rectangular channel above a resistance.
When the resistance is turned on, the PDMS is locally swollen, leading to mechanical deformation of the channel and drop, and the local fluids heat up, leading to local temperature gradients and corresponding Marangoni effects.
For weak deformations, we find that the droplet maintains its equilibrium steady shape, even under continuous heating. This equilibrium regime suggests that mechanical deformation dominates our system, with a negligible contribution from the thermal surface tension gradients. Accordingly, a second passive device that mimics the channel thickness variation induced by the thermal dilation of the channel walls in the thermomechanical device has the same dynamics as the active thermomechanical device, except that the constriction size and the fate of the droplet cannot be controlled. \\
We can therefore propose a minimal, semi-analytical model for the droplet dynamics and final state, focusing on mechanical effects and surface minimisation. We show that in both devices, the droplet's fate is determined by a global surface energy minimisation. The channel geometry and the extent of the confinement gradient set the neck size. If the neck size that minimises the surface energy is wider than the channel height, the droplet reaches an equilibrium steady-state configuration. If the neck is thinner than the channel height, the droplet breaks. 
We can therefore predict the droplet's fate for a given geometry using direct \textit{Surface Evolver} simulations or a minimal model that relies on the self-similarity of the neck shape.
The length of the gutters instead sets the dynamics of the deformation and breakup at the four channel corners. We model the time evolution of the deformation by balancing the rate of change in the droplet surface area with the viscous dissipation power in the gutters, using an adjustable parameter $K$ that accounts for the permeability in the gutters. With a single fitting parameter, our predictions are validated in both the thermomechanical and purely mechanical devices.\\
Conversely, our predictive model can quantify in situ out-of-plane topographic defects in a microfluidic channel theoretically down to a few hundred nanometers, which are challenging to measure with conventional microscopic or profilometric techniques. Furthermore, we believe that the thermomechanical device used for this study and inspired by previous work \cite{miralles2015versatile} can add a new, well-controlled, on-demand breakage function to the range of features explored in droplet microfluidics. 

\section{Acknowledgements}
This work was supported by Centre National de la Recherche Scientifique (CNRS), ESPCI Paris, Université de Rennes, and the Agence Nationale de la Recherche (ANR) under Grant No. ANR-18-CE09-0029, IPGG (Equipex ANR-10-
EQPX-34 and Labex ANR-10-LABX-31), PSL (Idex ANR-10-IDEX-0001-02). AT is grateful to the Simons Foundation through the Math + X grant awarded to the University of Pennsylvania, where part of the modelling has been performed.

\bibliographystyle{apsrev4-2}
\bibliography{biblio_MKbreakupdoi}

\vspace{30pt}

\setcounter{section}{0}
\appendix

\section{Increase of the interfacial tension with temperature measured by pendant drop method.}
\label{app:pendantdrop}
\begin{figure}[H]
\centering
\includegraphics[scale=0.6]{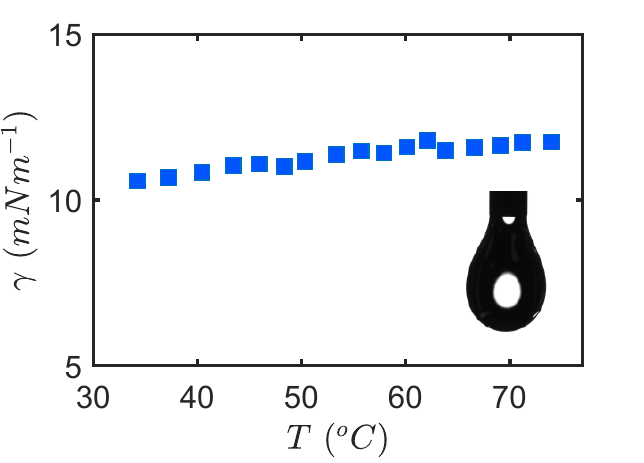}
\caption{ Interfacial tension as a function of the temperature, measured by the pendant drop method. Inset: picture of a drop of mineral oil in SDS solution.  }
\label{fig:ift}
\end{figure}
%
%
\section{Calibration device for the thermomechanical deformation height.}
\label{appendix:calibration}

\begin{figure}[H]
\centering
\includegraphics[scale=1.2]{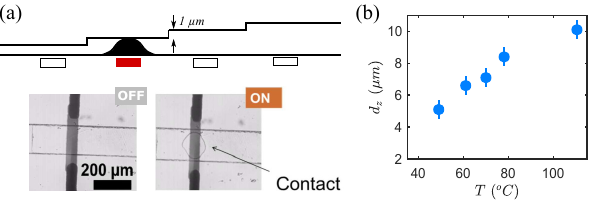}
\caption{ (a) Sketch of the $200 \mu$m wide channel with 1 $\upmu$m steps, obtained from micromilled brass mold and used to characterize the thermomechanical deformation. The contact zone is directly visible under the microscope. (b) Deformation height $d_z$ versus the maximum temperature of the heating resistance.  }
\label{fig:step}
\end{figure}

\section{Sensitivity of the inferred indentation size to the indentation profile.}
\label{appendix:indentation}

In this section, we discuss the droplet deformation's sensitivity to the channel indentation profile. Overall, we find that the predicted droplet deformation is only weakly sensitive to the indentation profile. We therefore expect our fitted values of ($d_x$,$d_z$) to be robust even if the indentation is not a linear ramp as in our model. 

\begin{figure}[hbt]
    \centering
    \includegraphics[width=0.95\linewidth]{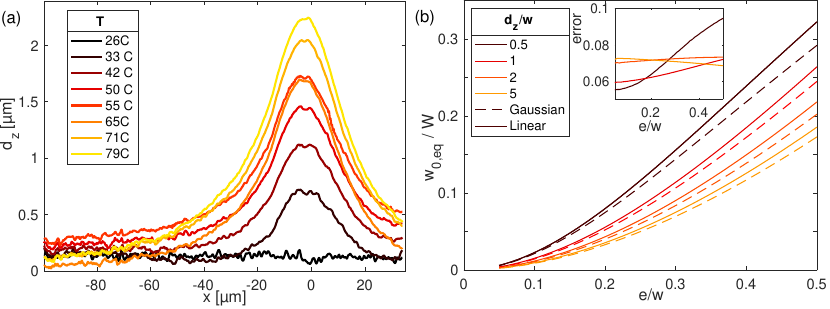}
    \caption{ Indentation shape in the channel. (a) Spatial profile of a layer of PDMS spin-coated on top of a heated resistance, which is an approximation for the thermomechanical device deformation. The profile is measured with a mechanical profilometer for different temperatures $T$ in the channel. (b) Difference between the equilibrium deformation $w_{0,eq}$ predicted by our static model for a linear ramp (plain line) and a Gaussian (dashed line) deformation profile with the same half-width, for different geometries. The inset shows the relative error between the two profiles.}
    \label{fig:Ngaussien}
\end{figure}

First, we measure the deformation of a layer of PDMS above a resistance to approximate the active deformation (Fig.~\ref{fig:Ngaussien}a). The profiles in Figure~\ref{fig:Ngaussien}(a) are obtained using a mechanical profilometer in an open system, that is, a single layer of 30 $\mu$m thickness of PDMS above the heating resistance, without the microfluidic system. This means that the real profile of the deformation, which is not possible to measure {\it in situ}, is probably lower than the one we measure. Nevertheless, we find that they are consistently curved and have an intermediate profile between a Gaussian and a linear ramp.

To understand the effect of the indentation shape on the droplet deformation, we then solve the model and compare the resulting necks for a linear and a Gaussian deformation with identical half-widths $d_x/2$ (Fig.~\ref{fig:Ngaussien}b). 
The difference between the corresponding equilibrium deformations is consistently less than $ 10\%$. For Figure 9 for example, instead of (130 $\mu$m, 6 $\mu$m) the Gaussian fit would be (122 $\mu$m, 6.59 $\mu$m). This shows that the droplet deformation depends only weakly on the indentation profile. This validates our approximation for the thermomechanical indentation as a ridge with a triangular cross-section. 

\section{Surface Evolver Simulations}
\label{appendix:codeSE}
We provide in this appendix the \textit{Surface Evolver} code for the simulations. 

To model the confined droplet in a channel, we start by implementing the confinement in the indented microfluidic channel as a list of global constraints that apply to all the vertices of the drop. Each represents one wall of the channel, including the plane indentation. We then define successively the nodes, oriented edges, oriented faces, and the body of the droplet. We finally provide examples of evolutions that can be used for surface minimization. \\
In the first implementation, the initial condition is an undeformed drop. Successive mesh refinements and surface minimization yield a deformed, peanut-like droplet that matches well the experiments.

\begin{adjustwidth}{2cm}{2cm}
\small{
\texttt{$//$ Parameters \\
parameter hh = 1    $//$ channel height \\
parameter ww = 4     $//$ channel width \\
parameter lg = 10   $//$ drop length \\
parameter dx = 2   $//$ indentation half length \\
parameter dz = .425   $//$ indentation height  \vspace{12pt} \\
$//$ --------------------------------------------------------- \\
$//$ Constraints: channel walls, all apply to all vertices \\
constraint 1  global  nonnegative  \\
	formula: z = 0  \\
constraint 2 global nonpositive  \\
	formula: z = hh \\
constraint 3 global  nonnegative \\
	formula: y = 0  \\
constraint 4 global  nonpositive \\
	formula: y = ww \\
constraint 5 global  nonnegative \\
	formula: z - dz*(1 - sqrt(x $\hat{\;}$ 2)/dx)  = 0 
 \vspace{12pt} \\
$//$ ---------------------------------------------------------  \\
$//$ Droplet vertices (block-shaped droplet with indentation over the entire droplet)
vertices  \\
1  -lg/2     0     0     constraint 1 3   \\
2  lg/2      0     0     constraint 2 3   \\
3  lg/2      ww    0     constraint 2 4  \\
4  -lg/2     ww    0     constraint 1 4  \\
5  -lg/2     0     hh    constraint 1 3  \\
6  lg/2      0     hh    constraint 2 3  \\
7  lg/2      ww    hh    constraint 2 4  \\
8  -lg/2     ww    hh    constraint 1 4   \\
9   0        0    dz   \\
10  0        ww     dz    \vspace{12pt} \\
$//$ ---------------------------------------------------------
$//$ Edges of the droplet
edges
1   1 9   \\ 
2   2 3   \\
3   3 10  \\
4   4 1 \\
5   5 6\\
6   6 7  \\
7   7 8 \\
8   8 5\\
9   1 5  \\ 
10  2 6  \\
11  3 7 \\
12  4 8\\
13  9 2 \\
14  10 4 \\
15 9 10  \vspace{12pt} \\
$//$ ---------------------------------------------------------
$//$ Faces of the droplet (correct cyclic ordering) \\
faces\\
1   1 13 10 -5  -9 \\
2   2 11 -6 -10  \\
3   3 14 12 -7 -11  \\
4   4  9 -8 -12 \\
5   5  6  7   8 \\
6  -4 -14 -15 -1  \\
7  15 -3 -2 -13  \vspace{12pt} \\
$//$ --------------------------------------------------------- \\
$//$ Body of the droplet \\
bodies \\
1  1 2 3 4 5 6 7  volume hh*ww*lg- dz*ww*lg/2  \vspace{12pt} \\
$//$ ---------------------------------------------------------  \\
read  \\
$//$ Example evolution \\
$//$ cycles with increasing degrees of mesh refinement  \\
refc := $\{$refine edges where length > .5; delete edges where length <.2$\}$	\\
cyc := $\{$refc 2; g 20; refc 2$\}$ \\
ref := $\{$refine edges where length > .3; delete edges where length <.1$\}$	\\
cy := $\{$ref 2; g 50; ref 2$\}$ \\
reff :=$\{$ refine edges where length > .15;   delete edges where length <.05$\}$ \\
cyf := $\{$reff 3; g 100; reff 2$\}$ \\	
$//$ proposed evolution pattern \\
$//$ note that convergence is very slow for this system, so longer runs can be needed depending on parameters \\
gogo := $\{$$\{$refc 2; g 3$\}$ 5;  cyc 10; cy 20; r; cyf; g 25 $\}$ \\
$//$ exportation of the vertices for shape extraction \\
expvert := $\{$foreach vertex do printf "$\%$f $\%$f $\%$f \textbackslash n",x,y,z >> "gouttever.txt" $\}$ }}
\end{adjustwidth}
\vspace{24pt}

 While the self-similar peanut shape of the deformation is very robust and reached quickly, the simulation then converges slowly to the final amplitude $w_{0,eq}$ because the area gradient descent is shallow. 
To ensure our results are robust, we developed a second implementation with different initial conditions. In this case, the droplet starts already deformed in the channel with an initial neck $w_i > w_{0,eq}$ . The neck thickens to reach its equilibrium shape as the surface is minimized. \\
For a given set of parameters, we consider that the system has converged when the two simulations have reached the same deformation amplitude $w_{0,eq}$. We then validate our theoretical static model by showing that the predicted deformation matches our simulations for different geometries. \\
\vspace{24pt}

\begin{adjustwidth}{2cm}{2cm}
\small{
\texttt{$//$ Parameters  \\
parameter hh = 1       $//$ channel height  \\
parameter ww = 4       $//$ channel width \\
parameter lg = 10      $//$ ~drop length \\
parameter dx = 2       $//$  indentation half length \\
parameter dz = .425     $//$  indentation height \\
parameter wi = 1.7     $//$  indentation height \vspace{12pt} \\
$//$ -------------------  \\
$//$ Droplet vertices (in this implementation, the droplet starts as a deformed -wi- block)
vertices  \\
1  -lg/2    0     0    \\
2  -dx      0     0    \\
3  0        wi    dz    \\
4  dx       0     0    \\
5  lg/2     0     0   \\
6  lg/2     0     hh    \\
7  dx       0     hh   \\
8  0        wi    hh   \\
9  -dx      0     hh   \\
10 -lg/2   0      hh   \\
11 -lg/2   ww     0    \\
12 -dx     ww     0    \\
13  0      ww-wi  dz    \\
14  dx     ww     0    \\
15  lg/2   ww     0   \\
16  lg/2   ww     hh    \\
17  dx     ww     hh   \\
18  0      ww-wi   hh   \\
19  -dx     ww     hh   \\
20  -lg/2   ww     hh   \vspace{12pt}\\
$//$ ------------------- \\
$//$ Edges of the droplet \\
edges \\
1   1 2 \\
2   2 3    \\
3   3 4 \\
4   4 5  \\
5   6 5 \\
6   6 7   \\
7   7 8 \\
8   8 9 \\
9   9 10 \\
10  10 1 \\
11   11 12 \\
12   12 13    \\
13   13 14 \\
14   14 15  \\
15   16 15 \\
16   16 17   \\
17   17 18 \\
18   18 19 \\
19   19 20 \\
20   20 11 \\
21 9 2 \\
22 8 3 \\
23 7 4 \\
24 19 12 \\
25 18 13 \\
26 17 14 \\
27 1 11 \\
28 2 12 \\
29 3 13 \\
30 4 14 \\
31 5 15 \\
32 6 16 \\
33 7 17 \\
34 8 18 \\
35 9 19 \\
36 10 20 \\
$//$ ------------------- \vspace{12pt} \\
$//$ Faces of the droplet (correct cyclic ordering) \\
faces \\
1   1 -21 9 10 \\
2   2 -22 8 21 \\
3   3 -23 7 22 \\
4   4 -5 6 23 \\
5   -11 -20 -19 24 \\
6   -12 -24 -18 25 \\
7   -13 -25 -17 26 \\
8   -14 -26 -16 15  \\
9   35 19 -36 -9 \\
10   34 18 -35 -8 \\
11   33 17 -34 -7 \\
12   32 16 -33 -6 \\
13   27 11 -28 -1 \\
14   28 12 -29 -2 \\
15   29 13 -30 -3 \\
16   30 14 -31 -4 \\
17   36 20 -27 -10  \\
18   -32 5 31 -15  \vspace{12pt} \\
$//$ -------------------  \\
$//$ Body of the droplet \\
bodies \\
1  1 2 3 4 5 6 7 8 9 10 11 12 13 14 15 16 17 18   volume  hh*ww*lg- dz*ww*lg/2 $//$ volume to match droplet.fe for convergence comparison  \\
$//$ hh*ww*lg-2*wi*dx*hh-2*(ww-wi)*dz*dx  }}
\end{adjustwidth}

\section{ Static model: droplet surface variation.}
\label{appendix:staticmod}

Here, we estimate the change in surface area that the drop undergoes in the presence of an indentation, that is, the difference between the surface energy of the peanut-shaped deformed droplet at time $t$ and that of the plug-like, undeformed one at $t = 0$. 

Because the droplet is centred at the level of the confinement gradient, the axis $x=0$ is an axis of symmetry for the droplet. Hence, we only consider the half-surface of the droplet, and we delineate three regions :

\begin{enumerate}[nolistsep,label={(\arabic*)}]
\item the droplet extremity, which only translates in the channel 
\item the sides of the droplet, which are elongated during the evolution, and where the gutters are localized 
\item the confinement gradient region where the neck forms. The extension of the neck in the $x$-direction is $b$, and its profile is $w(x,t)$.  
\end{enumerate}

\begin{figure}[ht]
\centering
\includegraphics[scale=1.2]{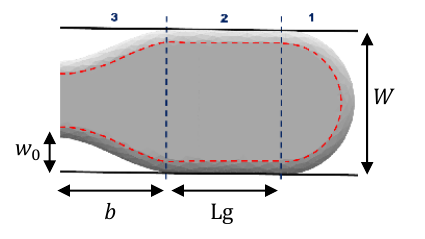}
\caption{Regions 1, 2, and 3 for the detailed calculation of $\Delta S$.} 
\label{fig:Delta_S}
\end{figure}

Figure \ref{fig:Delta_S} illustrates regions 1, 2, and 3. We calculate the excess surface of the droplet interface in each of these regions. 

\paragraph{Region 1:}
The surface of the droplet interface is not modified in this region, which translates into the channel without changing shape.

\paragraph{Region 2:}
In region (2), the confinement height is constant and equal to $e$. 
By mass conservation, as the neck forms, the length of region (2), $l(t)$, therefore increases by $\Delta L/2$, with $\Delta L$ the total length increase of the drop. 

The total surface change of the region (2), including the menisci, is given by
\begin{equation}
 \Delta S_2= \left[2(W-e)+ 2 \pi \frac{e}{2}\right] \Delta L/2, 
\label{surfaceini}
\end{equation}
while its volume change is 
\begin{equation}
\Delta \Omega_{2}= \left[(W-e) e + \pi \left(\frac{ e}{2}\right)^2\right]\Delta L/2. 
\end{equation}

\paragraph{Region 3:}
The area of region 3 is more intricate to evaluate, as it has a complex shape, set by the confinement height $e(x)$ and the in-plane distance from the neck to the wall $w(x,t)$. We can, however, evaluate it by recalling that the slope of the indentation is small, $d_z \ll d_x$. 
We consider separately the changes in surface area of the top surface, $\Delta S_3^t$, the bottom one in contact with the indentation, $\Delta S_3^b$, and the menisci $\Delta S_3^m$. 

The top and bottom surfaces are flat. The change in the surface area of the top surface is
\begin{equation}
\begin{split}
\Delta S_3^t &=\left[W b - 2\int^{{b}}_{0}\left(w(x,t) + \frac{e(x)}{2}\right)\textrm{d} x\right] - \left[ W -e (x) \right]b \\
 & = - \int^{{b}}_{0} 2 w(x,t)  \textrm{d} x, 
\end{split}
\end{equation}
while that of the bottom one which is tilted along the indentation, is
\begin{equation}
\begin{split}
\Delta S_3^b(t) = & - 2 \int^{{b}}_{0}  w(x,t) \sqrt{1 + \left(\partial _x e \right)^2}  \textrm{d} x .
\end{split}
\end{equation}

The local channel height $e(x)$ and the constant mean curvature of the entire drop set the shape of the two menisci. In the limit of small indentations, $d_z \ll d_x$, we can approximate it for a given neck shape $w(x,t)$ by considering small cylindrical sections of radius $e(x)/2$ and height $\textrm{d} s$ along the arclength $s$ of the meniscus. Noting that $\textrm{d} s = \textrm{d}x \sqrt{1 + (\partial_x w) ^2 }$, we get the surface area change 
\begin{equation}
\begin{split}
\Delta S_3^m =2 \int^{{b}}_{0} \pi e(x)(\sqrt{1+\partial_x w^2}-1)\textrm{d} x. 
\end{split}
\end{equation}
Finally, the total area change in region (3) is 
\begin{equation}
\begin{split}
\Delta S_3 = - 2 \int^{{b}}_{0} \left( 1+  \sqrt{1 + \left(\partial _x e \right)^2} \right) w(x,t)  \textrm{d} x + 2 \int^{{b}}_{0} \pi e(x)\left(\sqrt{1+\partial_x w^2}-1 \right)\textrm{d} x. 
\end{split}
\end{equation}

Similarly, we obtain the volume change 
\begin{equation}
\begin{split}
\Delta \Omega_3 = - 2 \int^{{b}}_{0}  w(x,t) e(x) \textrm{d} x + \frac{\pi}{4} \int^{{b}}_{0}  e(x)^2 \left(\sqrt{1+\partial_x w^2}-1 \right)\textrm{d} x. 
\end{split}
\label{eqneckvolume}
\end{equation}

\paragraph{Mass conservation.}By volume conservation of the droplet, $\Delta\Omega_2 = \Delta \Omega _3$, we obtain the elongation of the droplet $\Delta L$ as 
\begin{equation}
    \Delta L  = 2 \frac{1}{W + \left(\frac{\pi}{4} - 1 \right)e}  \left[   - 2 \int^{{b}}_{0}  w(x,t) \frac{e(x)}{e} \textrm{d} x + \frac{\pi}{4} \int^{{b}}_{0}  \frac{e(x)^2}{e} \left(\sqrt{1+\partial_x w^2}-1 \right)\textrm{d} x \right] . 
    \label{eq:lengthchange}
\end{equation}

The total surface change of the entire droplet, $\Delta S$, is then 
\begin{equation}
\begin{split}
\Delta S = 4 \int^{{b}}_{0} \left\{\pi e(x)\left(\sqrt{1+\partial_x w^2}-1 \right) \left[ 1 + \frac{\lambda}{4} \frac{e(x)}{e} \right]   -   w(x,t) \left[ 1+ \sqrt{1 + \left(\partial _x e \right)^2}   + 2 \lambda \frac{e(x)}{e} \right] \right\} \textrm{d} x  ,
\end{split}
\end{equation}
where 
\begin{equation}
    \lambda = \frac{W + \left(\frac{\pi}{2}-1 \right)e }{W + \left(\frac{\pi}{4}-1 \right)e}
\end{equation}
is a geometric parameter of the unindented channel with $1 < \lambda < 2$. 

We now use the result from our experiments and \textit{Surface Evolver} simulations that the neck profile is self-similar and can be collapsed on a single function $p(u)$ with $u \in [-1, 1]$.
Note that up to now, our calculations make no hypothesis on the shape of the indentation, which is only described by the changing height $e(x)$. We now consider a triangular indentation, as for the mechanical device.  
We can then write that for $-b<x<b$, $x = b u $, $w(x,t) = w_0(t) p(u)$ and $e(x) =  e - d_z  + d_z \min(1, \frac{b}{d_x}|u|) $. Denoting for conciseness $\bar u = \min(1, \frac{b}{d_x}|u|)$, 

\begin{equation}
\begin{split}
\Delta S (w_0, b) = 4 b \int^{{1}}_{0} & \left\{  \pi \left(\sqrt{1+  \left(\frac{w_0}{b} \right)^2 p'(u)^2}-1 \right)  \left(e - d_z (1-\bar u)\right)  \left[ 1 + \frac{\lambda}{4} \left(1 - \frac{d_z}{e} (1-\bar u) \right) \right] \right.  \\ 
 & \left. -  w_0  p(u) \left[ 1+  \sqrt{1 + \left(\partial _x e \right)^2}  + 2 \lambda \left(1 - \frac{d_z}{e} (1-\bar u) \right) \right] \right\} \textrm{d} u  . 
\end{split}
\label{eq:surfacess}
\end{equation}
The surface area is set by the amplitude of the neck $w_0$, its aspect ratio $w_0/b$, and the geometry of the channel ($W$, $e$) and the indentation ($d_x,d_z)$.
The equilibrium neck shape ($w_{eq},b_{eq}$) corresponds to the minimum of $\Delta S$, which we then obtain numerically using the built-in Matlab function \textit{fminsearch}.  

\begin{figure}[ht]
\centering
\includegraphics[scale=0.5]{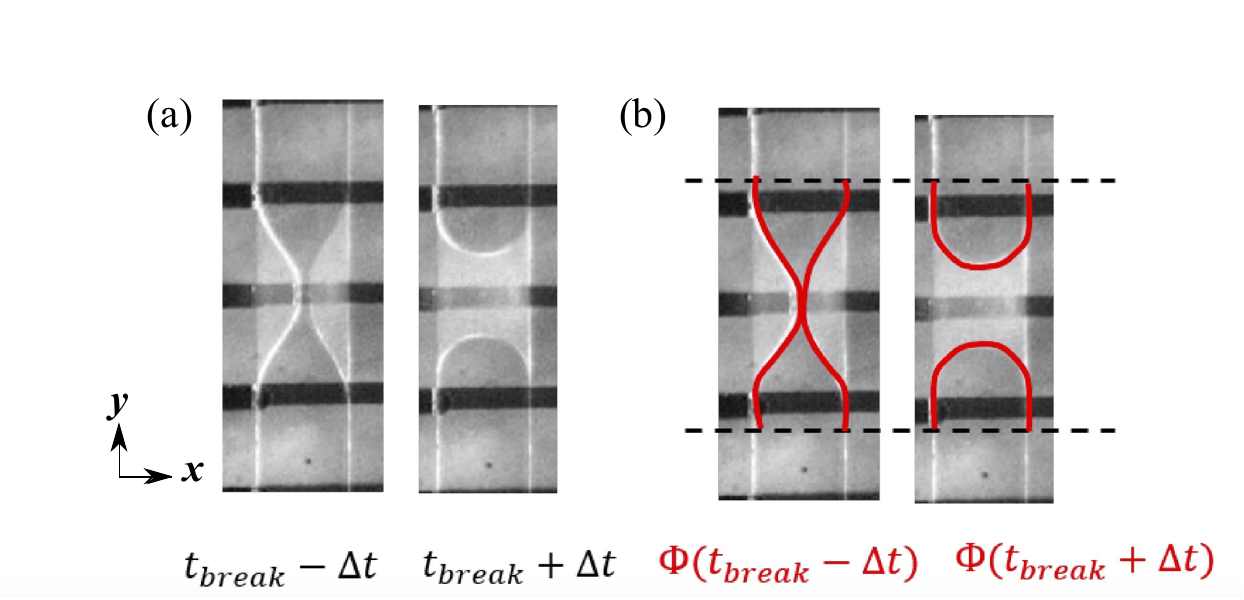}
\caption{Calculation of the droplet in-plane interface perimeter at the level of the neck, just before and just after the breakup event ($\Phi(\tau_{break}-\Delta t)=512$ $\upmu$m and that
$\Phi(\tau_{break}+\Delta t)=460$ $\upmu$m).} 
\label{fig:perimetre}
\end{figure}

\section{Polynomial approximation of the self-similarity function for the neck profile}
\label{appendix:pfunction}
The function $p$ is extracted from the experimental data. More precisely, $p$ is obtained by fitting the experimental profiles illustrated in FIG.~\ref{fig:selfsimilar}.
We find for $0<u<1$,
\begin{equation}
p(u)=-1.6727u^5+3.3531u^4+0.317u^3-3.0016u^2+0.0033u+1.0001
\label{p}
\end{equation}

\section{Dynamic model: time-evolution of the neck.}
\label{appendix:dynamicmod}

\subsection{Scaling approach to the force balance, small deformations}
The neck formation is driven by surface energy minimization of the drop, which is instantaneously compensated by the viscous dissipation from the generated flows.

We derive a scaling law for the force balance when the neck forms.
At the deformation, the capillary pressure is $P_\gamma \sim \gamma \mathcal{C}$ with the local curvature $\mathcal{C}=\displaystyle\frac{\partial_x^2 w}{\left(1+(\partial_x w)^2 \right)^{3/2}}+\frac{2}{e(x)}
$.
The driving capillary force per unit length scales as $F_{\gamma} \sim \partial_x P_\gamma$. In the limit of small deformation at the neck, $\partial_x w \ll 1$, we obtain $F_{\gamma} \sim \gamma e^2 \partial_x^3 w$, while the viscous force per unit length writes: $F_{visc} \sim \eta e L_g (v_g/e^2)$, with $v_g$ the typical flow velocity in the gutters. 
This gutter velocity $v_g$ can be estimated by mass conservation, writing that the flow rate in the gutters is equal to the change in volume $\Omega$ at the neck $\partial_t \Omega \sim v_g e^2$.
Since $\partial_t \Omega \sim e b \partial_t w$, we get $v_g \sim \partial_t w (b/e)$ and $F_{visc} \sim \eta_w (\partial_t w) (b L_g/e^2)$. Then, a simple scaling law balancing the capillary driving force $F_{\gamma}$ and the viscous dissipation $F_{visc}$ yields: 
\begin{equation}
 \gamma e^2 \frac{\partial^3 w}{\partial x^3} \sim - \eta_w  \frac{b L_g}{e^2} \frac{\partial w}{\partial t}.
\end{equation}
Injecting the self-similar form of $w(x,t)=w_0(t)p(x/b_{eq})$, we obtain:
\begin{equation}
 \frac{\gamma e^2}{b_{eq}^3} \frac{\partial^3 p(u)}{ \partial u^3} w_0 \sim - \eta_w  \frac{b_{eq} L_g}{e^2} p(u) \frac{\partial w_0}{\partial t}
\end{equation}
which gives:
\begin{equation}
 \frac{1}{w_0}\frac{\partial w_0}{\partial t} \sim - \frac{\gamma e^4}{\eta_w b_{eq}^4 L_g} \frac{1}{p(u)}\frac{\partial^3 p(u)}{\partial u^3}
 \label{eq:forcebalance}
\end{equation}

\begin{figure}
    \centering
    \includegraphics[width=0.8\linewidth]{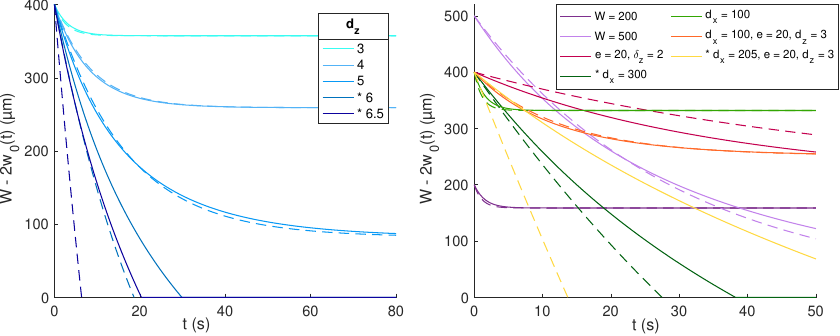}
    \caption{Comparison between our dynamic model (full line) and the exponential approximation in eq.~\eqref{eq:forcebalance} (dashed lines). We vary the channel geometry as indicated in the legend. When not stated otherwise, the parameters are $W = 400 \,\mu$m, $e = 30 \,\mu$m, $d_z = 5 \,\mu$m and $d_x = 200 \mu$m. Cases denoted by $*$ are breaking droplets, where the scaling model does not match the full predictions. }
    \label{fig:scalingdyn}
\end{figure}

This suggests that the neck exponentially relaxes to its equilibrium state with a characteristic timescale $ \sim \frac{\eta_w b_{eq}^4 L_g}{\gamma e^4}$, and confirms that the dynamics scale linearly with the gutter length $L_g$. 
This approach matches experiments for small indentations that lead to an equilibrium deformation. 
However, in the regime of larger indentation (typically see for $d_x= 205$ $\mu$m, in yellow, in figure  \ref{fig:scalingdyn}), when droplets eventually break up, this approach predicts arbitrarily fast, unphysical dynamics that do not match our experiments. We therefore move to a more quantitative approach that accounts for the time dynamics of the neck formation, building on our above static model. 

\subsection{Dynamic power balance}

To model the droplet deformation dynamics, we write a power balance that compares the viscous dissipation power with the excess surface energy in the system per unit time.

In our system, viscous dissipation occurs in the gutters of the drop and the neck region. As the gutters are highly confined, we expect them to dominate the viscous dissipation, which is confirmed by showing that the breakup time in a given geometry scales linearly with the droplet length.
We therefore only balance the power from surface area minimization $\mathcal{P}_\Sigma$ and the viscous dissipation in the gutters $\mathcal{P}_g$. 

\paragraph{Viscous dissipation in the gutters}

The viscous dissipation power for a flow with velocity $\mathbf{v}$ in a volume $\Omega$ is given by
\begin{equation}
\mathcal{P}=\oint _\Omega \eta (\nabla v+{}^T\nabla v)^2 \mathrm{d} \mathbf{\Omega}
\label{eq:gutterdiss}
\end{equation}
where $\eta$ is the viscosity of the fluid.
In the gutters, this power scales as 
\begin{equation}
\mathcal{P}_g\sim\eta_w \frac{v^2}{e^2}e^2L_g
\label{power2}
\end{equation}
where $v$ is the characteristic velocity in the gutters. 
By conservation of mass, the flow rate of water that fills the neck region (3) as the drop deform is equal to the flow rate in the gutters, that is, $ve^2 \sim \partial_t\Omega_3$, with $\Omega_3$ given in \eqref{eqneckvolume}. 

The length of the gutters $L_g(t)$ changes in time as the droplet deforms, as described in \eqref{eq:lengthchange}. The drop elongation is bounded, $\Delta L < b, w_0, W/2$. In the following, we work in the limit of long drops, $L \gg W$, we can instead assume that $L_g$ is constant in time $L_g(t) = L_g$.  
We also introduce a constant dimensionless coefficient $K$ that accounts for the flow profile in the gutters, independently of the channel geometry (see Appendix~\ref{appendix:flowgutter}), and write 
\begin{equation}
\mathcal{P}_g=K\frac{\eta_wL_g}{e^4}\left[\partial_t\Omega_3\right]^2. 
\label{powerf}
\end{equation}
Using the self-similar form of the neck profile $w(x,t) = w_0(t) p(x/b)$, with $b$ constant over time as observed experimentally,
\begin{equation}
\begin{split}
\partial _t\Omega_3 = \partial_t w_0(t) \int^{1}_{0}  \left[- 2 b  p(u)  e(bu) + \frac{\pi}{4}  \frac{w_0(t)}{b} e(bu)^2  p'(u)^2  \left(1+ \frac{w_0(t)^2}{b^2} p'(u)^2 \right) ^{-3/2}   \right]\textrm{d} u. 
\end{split}
\end{equation}
Therefore, the viscous dissipation power writes
\begin{equation}
\mathcal{P}_g=K\frac{\eta_wL_g}{ e^4} (\partial_t w_0)^2 \mathcal{I}(w_0,b),
\label{powerf3}
\end{equation}
where the integral $\mathcal{I}$ depends on the neck extension $(w_0,b)$ and the channel and indentation geometry, 
\begin{equation}
    \mathcal{I} (w_0,b) = \left[ \int^{1}_{0}  \left(  - 2 b  p(u)  e(bu) + \frac{\pi}{4}  \frac{w_0(t)}{b} e(bu)^2  p'(u)^2  \left(1+ \frac{w_0(t)^2}{b^2}  p'(u)^2 \right) ^{-3/2}   \right)\textrm{d} u \right]^2. 
\end{equation}

\paragraph{Power from the surface energy minimization}
The power from the decrease in surface energy is $\mathcal{P}_\Sigma=\gamma \partial_t(\Delta S)$. We obtain the rate of change of the surface area of the drop by deriving eq.~\eqref{eq:surfacess} with respect to time, 
\begin{equation}
\begin{split}
\partial_t \Delta S   =   \partial_t w_0  \int^{{1}}_{0} & 4 \left\{  \pi  \frac{w_0(t)}{b} p'(u)^2  \left(1+ \frac{w_0(t)^2}{b^2}  p'(u)^2 \right) ^{-3/2}  e(bu)  \left[ 1 + \frac{\lambda}{4} \frac{e(bu)}{e}\right] \right.  \\ 
 & \left. -  b p(u) \left[ 1+ \sqrt{1+ (\partial_x e)^2}   + 2 \lambda \frac{e(bu)}{e} \right] \right\} \textrm{d} u  \\
   = \partial_t w_0  \mathcal{J}(w_0,b), 
\end{split}
\end{equation}
where the integral $\mathcal{J}_0(w_0,b)$ is a function of the neck size $(w_0,b)$ and depends on the channel geometry. 
Similarly to the case of the viscous dissipation, we therefore write the power from surface energy minimization as 
\begin{equation}
    \mathcal{P}_\Sigma=  \gamma \partial_t w_0 \mathcal{J}(w_0,b). 
\end{equation}

\paragraph{Power balance}
We now use that the extension of the neck $b$ in the $x$-direction is constant, $b = b_{eq}$, and can be computed from the static model above.
From the power balance $ \mathcal{P}_\Sigma = \mathcal{P}_g$, we therefore get that 
\begin{equation}
\begin{split}
 \partial_t w_0 = \frac{\gamma e^4}{K\eta_wL_g} \frac{\mathcal{J}(w_0 ,b_{eq})}{\mathcal{I}(w_0 ,b_{eq})}.
\label{balance2}
\end{split}
\end{equation}
We can solve this equation numerically using a forward Euler method, starting from the undeformed droplet $w_0(0) = 0$ as an initial condition. 
The  dimensionless adjustable parameter for the gutter flow $K$ is set by fitting this prediction for the neck dynamics $w_0(t)$ to the experimental data from the mechanical devices, in which the geometric parameters ($d_x$, $d_z$, $W$ and $e$) are well characterized, and obtain that $K=4.68\times 10^{3}$. \\

\paragraph{Asymptotic limit in the breaking regime.}

When the droplets eventually break up, that is when the equilibrium deformation $w_{0,eq}$ is large ($w_{0,eq} \gg b, W$), we find that the rate of deformation above becomes independent of the neck size $w_{0}$, 
\begin{equation}
\begin{split}
 \partial_t w_0 = \frac{\gamma e^4}{K\eta_w L_g b_{eq}} \frac{  \int^{{1}}_{0}   p(u) \left[ 1+ \sqrt{1+ (\partial_x e)^2}  + 2 \lambda \frac{e(bu)}{e} \right] \textrm{d} u }{ \left[ \int^{1}_{0} p(u)  e(bu)  \textrm{d} u \right]^2 }.
\end{split}
\end{equation}
As a result, the deformation occurs linearly for breaking droplets, as seen in experiments.

\section{Flow profile in the gutters.}
\label{appendix:flowgutter}

Our dynamic model in Appendix~\ref{appendix:dynamicmod} relies on using a fitted constant $K$ accounting for the flow in the droplet gutters. We show the agreement between our model and experiments by independently estimating this constant in \textsc{Comsol} finite element simulations. We consider two limiting cases of boundary conditions at the liquid-liquid interface: either a no-slip condition (interface velocity equals zero) or a stress-free boundary condition ($\partial_\tau v=0$, {\it i.e.} no tangential stress at the interface). The flow at the channel wall is always no-slip. 
We obtain the value of the constant $K$ in our simulations as detailed below, and find $K_{ns} \approx 6 \times 10^3$ with a no-slip condition and $K_{sf} \approx 2.8 \times 10^3$ with a stress-free condition. Strikingly, the experimental value for the fit to our dynamical model is $K=4.68\times 10^{3}$, between these two limit theoretical values, further validating our theoretical approach. 

More precisely, we run 3D stationary simulations of the Stokes equations in the gutter geometry driven by a constant pressure $p_0$. The laminar flow is in the $x$-direction, and we consider a cross-section away from the inlet and outlet so that the flow is independent of $x$, $\mathbf{u} = u(y,z)$. The resulting velocity depends on the box length and the arbitrary $p_0$, but not the velocity-to-shear ratio we are ultimately interested in. 

The viscous dissipation in a gutter of length $L_g$ and cross-section $S$ is 
\begin{equation}
\begin{split}
    \mathcal{P}_g & = L_g \oint _S \eta (\nabla v+{}^T\nabla v)^2 \mathrm{d} S \\
      & =  2 L_g \eta \oint _S \dot \gamma ^2 (y,z) \mathrm{d} S \\
      & = 2 L_g \eta S \langle \dot \gamma ^2 \rangle_S 
\end{split}
\end{equation}
with $\dot \gamma$ the local shear rate and, $\langle ... \rangle_S$ denoting an average value over the cross section. 

The volumetric flow rate in the four left gutters is equal to the rate of change of the volume in the half region of the neck, that is 
\begin{equation}
    \begin{split}
        \partial_t\Omega_3  &=  \oint _S  u(y,z) \mathrm{d} S  \\
                            & = S \langle u \rangle_S.
    \end{split} .
\end{equation}

Introducing the constant $K_{th}$ as in eq.~\eqref{powerf} to account for the flow in the wedge, we get that the two above expressions for the viscous dissipation power are equal when 
\begin{equation}
    2 L_g \eta S \langle \dot \gamma ^2 \rangle_S = K\frac{\eta_wL_g}{e^4} S^2 \langle u \rangle_S ^2.
\end{equation}
 Using the cross section for the four half gutters of length $L_g$, $S = e^2 - \pi (e/2)^2$, we get the following expression for $K$:
\begin{equation}
    K = \frac{\langle \dot \gamma ^2 \rangle_S}{\langle u \rangle_S ^2} \frac{2 e^2}{1 - \pi/4}.
\end{equation}
We then use these expressions to obtain the above theoretical values of $K$ in gutters. \\

\begin{figure}[h]
    \centering
    \includegraphics[width=0.85\linewidth]{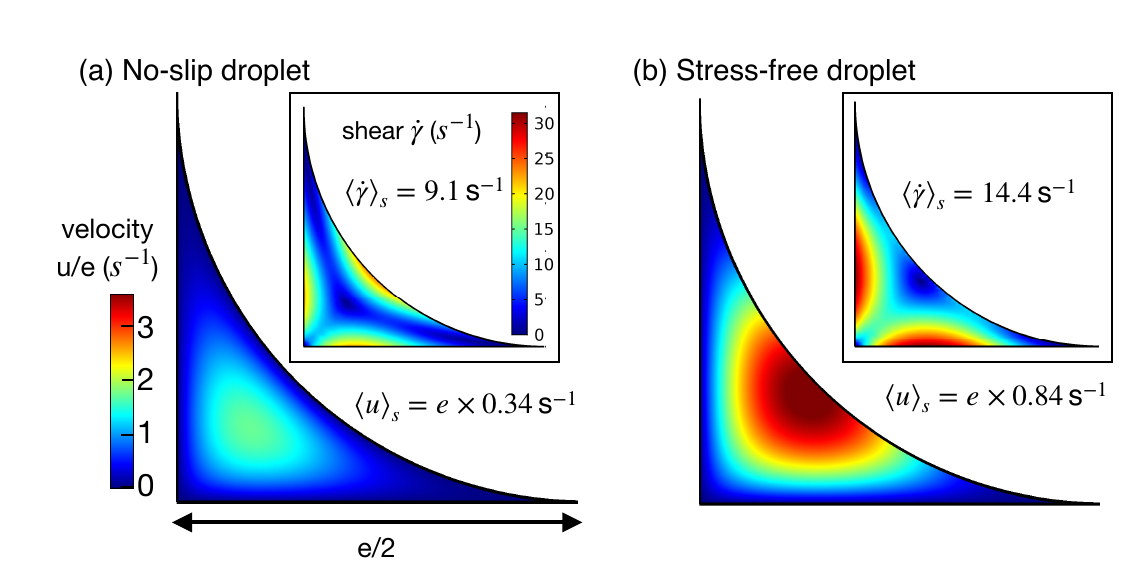}
    \caption{flow in a gutter cross-section with a (a) no-slip and (b) stress-free boundary condition at the liquid interface. The main figures show the velocity, and the insets show the shear rate and their average value over the cross-section (both in arbitrary units). }
    \label{fig:enter-label}
\end{figure}

\section{Sensitivity of the error function $N$ to small variations near the best fitting values of ($d_x$,$d_z$).}
\label{appendix:nsensitivity}
\begin{figure}[H]
\centering
\includegraphics[scale=.8]{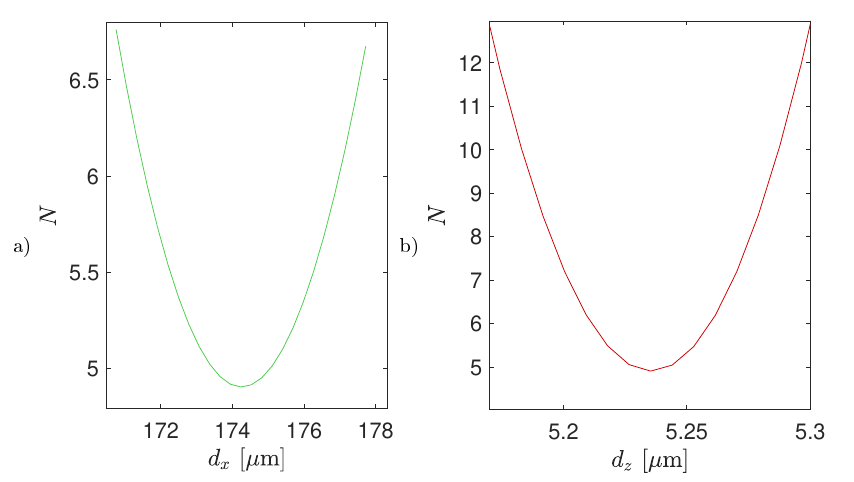}
\caption{Evolution of the  error function $N(d_x,d_z)=4[w_0^{exp}(t)-w_0^{th}(t)]^2/e_{rr}^2$ as a function of $d_x$ (a) and $d_z$ (b) small variations near the best fitting values ($d_x$, $d_z$) shown in Fig.\ref{fig:modelinverse}.}
\label{fig:error}
\end{figure}

\end{document}